\documentclass[aps,pra,twocolumn,superscriptaddress,footinbib,amsmath,amssymb]{revtex4}
\usepackage{amsmath}
\usepackage{gensymb}
\usepackage{dcolumn}
\usepackage{bm}
\usepackage[pdftex]{graphicx,color} 
\usepackage{epstopdf} 
\usepackage{graphicx}
\usepackage{float}
\usepackage[english]{babel}
\usepackage[normalem]{ulem} 

\usepackage{array}
\newcolumntype{P}[1]{>{\centering\arraybackslash}p{#1}}
\newcolumntype{M}[1]{>{\centering\arraybackslash}m{#1}}

\begin{document}
\title{Three-dimensional photonic band gap cavity with finite support: enhanced energy density and optical absorption}

\author{D. Devashish}
\affiliation{Complex Photonic Systems (COPS), MESA+ Institute for Nanotechnology, University of Twente, P.O. Box 217, 7500 AE Enschede, The Netherlands}
\affiliation{Mathematics of Computational Science (MACS), MESA+ Institute for Nanotechnology, University of Twente, P.O. Box 217, 7500 AE Enschede, The Netherlands}
\affiliation{Present address: ASML Netherlands B.V., 5504 DR Veldhoven, The Netherlands}

\author{Oluwafemi Ojambati}
\affiliation{Complex Photonic Systems (COPS), MESA+ Institute for Nanotechnology, University of Twente, P.O. Box 217, 7500 AE Enschede, The Netherlands}
\affiliation{Present address: Cavendish Laboratory, Department of Physics, NanoPhotonics Centre, University of Cambridge, Cambridge CB3 0HE, United Kingdom}

\author{Shakeeb B. Hasan}
\affiliation{Complex Photonic Systems (COPS), MESA+ Institute for Nanotechnology, University of Twente, P.O. Box 217, 7500 AE Enschede, The Netherlands}
\affiliation{Present address: ASML Netherlands B.V., 5504 DR Veldhoven, The Netherlands}

\author{J. J. W. van der Vegt}
\affiliation{Mathematics of Computational Science (MACS), MESA+ Institute for Nanotechnology, University of Twente, P.O. Box 217, 7500 AE Enschede, The Netherlands}

\author{Willem L. Vos}
\email{w.l.vos@utwente.nl}
\affiliation{Complex Photonic Systems (COPS), MESA+ Institute for Nanotechnology, University of Twente, P.O. Box 217, 7500 AE Enschede, The Netherlands}
\homepage{www.photonicbandgaps.com}

\date{August 16th, 2018}

\begin{abstract}
We study numerically the transport and storage of light in a three-dimensional (3D) photonic band gap crystal doped by a single embedded resonant cavity.
The crystal has finite support since it is surrounded by vacuum, as in experiments. 
Therefore, we employ the finite element method to model the diamond-like inverse woodpile crystal that consists of two orthogonal arrays of pores in a high-index dielectric such as silicon and that has experimentally been realized by CMOS-compatible methods. 
A point defect that functions as the resonant cavity is formed in the proximal region of two selected orthogonal pores with a radius smaller than the ones in the bulk of the crystal. 
We present a field-field cross-correlation method to identify resonances in the finite-support crystal with defect states that appear in the 3D photonic band gap of the infinite crystal. 
Out of five observed angle-independent cavity resonances, one is $s$-polarized and four are $p$-polarized for light incident in the $X$ or $Z$ directions.  
It is remarkable that quality factors up to $Q = 1000$ appear in such thin structures (only three unit cells), which is attributed to the relatively small Bragg length of the perfect crystal. 
We find that the optical energy density is remarkably enhanced at the cavity resonances by up to $2400 \times$ the incident energy density in vacuum or up to $1200 \times$ the energy density of the equivalent effective medium. 
We find that an inverse woodpile photonic band gap cavity with a suitably adapted lattice parameter reveals substantial absorption in the visible range. 
Below the 3D photonic band gap, Fano resonances arise due to interference between the discrete fundamental cavity mode and the continuum light scattered by the photonic crystal. 
We argue that the five eigenstates of our 3D photonic band gap cavity have quadrupolar symmetry, in analogy to d-like orbitals of transition metals.  
We conclude that inverse woodpile cavities offer interesting perspectives for applications in optical sensing and photovoltaics. 
\end{abstract}

\maketitle


\section{Introduction}
Confining light in a minuscule volume in space is a main topic in nanophotonics~\cite{Vahala2003Nature, Lourtioz2005Book, Novotny2006Book,Ghulinyan2015Book}. 
Notable interests include trapping or slowing down of photons~\cite{Baba2008NatPhot}, sensing for bio applications~\cite{Krioukov2002OptLett}, Purcell enhancement of spontaneous emission~\cite{Purcell1961PR, Gerard1998PRL}, and cavity quantum electrodynamics~\cite{Thompson1992PRL, Imamoglu1997PRL, Gerard2003TAP, Reithmaier2004Nature, Yoshie2004Nature, Peter2005PRL}. 
To achieve micro- and nano-scale light confinement, many types of devices have been reported such as microspheres~\cite{Gorodetsky1996OptLett, Vernooy1998OptLett}, micropillars~\cite{Gerard1998PRL, Reithmaier2004Nature, Peter2005PRL, Pelton2002IEEEJQE}, microdisks~\cite{Gayral1999APL}, plasmonic cavities~\cite{Miyazaki2006PRL, Kuttge2010NanoLett, Chikkaraddy2016Nature}, toroidal rings~\cite{Armani2003Nature}, or 2D photonic crystal slab cavities~\cite{Painter1999Science, Akahane2003Nature}.   

A cavity formed by a defect embedded in a three-dimensional (3D) photonic band gap crystal has a prime significance~\cite{Yablonovitch1987PRL, John1987PRL}, since the confinement of light is truly three-dimensional~\cite{Busch2006book,Minkov2017APL}. 
A defect in a photonic crystal is formed by the addition or by the removal of high-index material to break the periodic spatial symmetry~\cite{Villeneuve1996PRB}. 
A defect formed by adding high-index material is called a donor defect since the defect state derives from the high frequency (``conduction") bands, whereas a defect formed by removing high-index material is called an acceptor defect since the defect state derives from the low frequency (``valence") bands~\cite{Villeneuve1996PRB, Ozbay1995PRB, Okano2002PRB,Joannopoulos2008Book}. 

Among all 3D photonic band gap crystals, the class of crystals with a diamond-like symmetry stand out for their broad 3D photonic band gap~\cite{Maldovan2004NP}, which makes them robust to unavoidable fabrication disorder~\cite{Woldering2009JAP}, and which is favorable to shield embedded cavities from the surrounding vacuum~\cite{Vos2015Book}. 
Following the seminal microwave studies of Yablonovitch \textit{et al.}~\cite{Yablonovitch1991PRL} and Bayindir \textit{et al.}~\cite{Bayindir2000PRL,Bayindir2000PRB}, Ogawa \textit{et al.} studied the modified emission spectra of quantum wells in presence of a 3D photonic band gap cavity in direct woodpile crystals made from GaAs~\cite{Ogawa2004Science}. 
The Kyoto group performed a group-theoretical study to identify the symmetry properties of the sizable number of cavity resonances that appear in woodpile crystals (between 17 and 32, depending on the cavity geometry)~\cite{Okano2004PRB}. 
Recently, the Tokyo team demonstrated a quality factor $Q = 12800$ in emission spectra of quantum dots in a GaAs woodpile crystal~\cite{Tajiri2015APL}. 

\begin{figure}[tbp!]
\centering
\includegraphics[width=0.75\columnwidth]{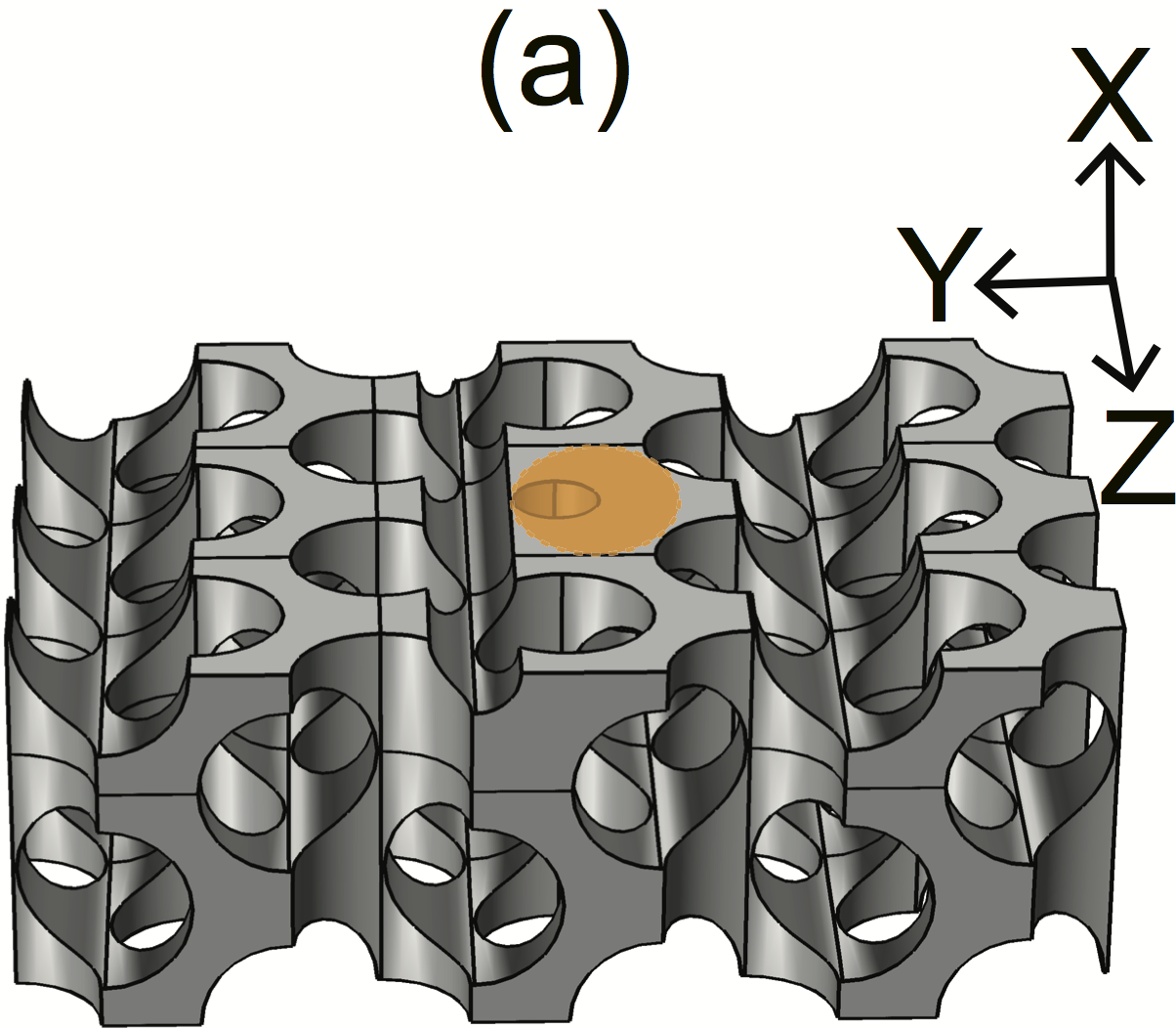} 
\includegraphics[width=0.9\columnwidth]{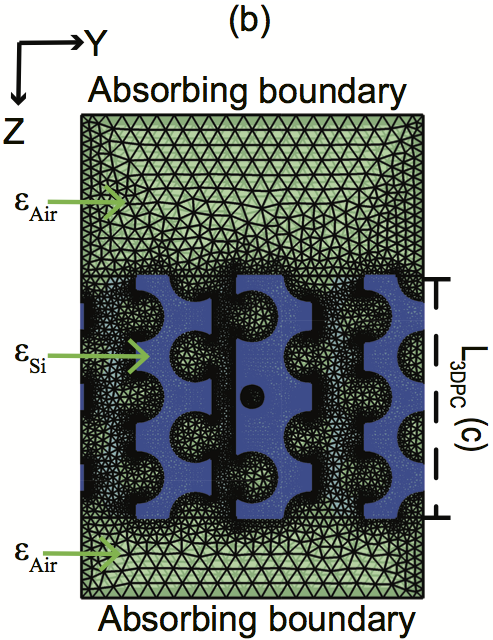}
\caption{(a) Schematic view of a $3 \times 3 \times 3$ supercell of a cubic inverse woodpile photonic crystal with a point defect, as a $YZ$ cut through the center of the cavity. 
The proximal region of the two defect pores with a radius smaller than all other pores has excess high-index material, as indicated by the orange ellipse, that functions as a 3D photonic band gap cavity.
(b) $YZ$ cross-section of the computational cell through the center of the cavity, showing the discretization mesh.
The cell with thickness $L_{3DPC} = 3c$ is bounded by absorbing boundaries at $-Z$ and $+Z$, and by periodic boundary conditions at $\pm X$ and $\pm Y$. 
The blue color represents the high-index backbone (silicon), and the green color represents air.
} 
\label{fig:CavityStructure}
\end{figure}

Here, we pursue the so-called inverse woodpile photonic crystal structures that consists of two 2D orthogonal arrays of pores~\cite{Ho1994SSC}. 
Since this structure is relatively straightforward to define, it has been realized by various nanofabrication techniques and high-index backbones~\cite{Schilling2005APL,Santamaria2007AdvMater,Hermatschweiler2007AdvFunctMater,Jia2007JApplPhys}. 
Moreover, 3D inverse woodpile photonic crystals made from silicon have been fabricated by our group using CMOS-compatible nanofabrication methods that were developed in collaboration with high-tech industry~\cite{Tjerkstra2011JVSTB,vandenBroek2012AFM,Grishina2015Nanotechnology}. 
To functionalize inverse woodpile crystals, our group has proposed a design to create a resonant cavity in the 3D inverse woodpile crystal structure, whereby the photons are tightly confined in the proximal region of two orthogonal defect pores that have a radius smaller than all other pores in the bulk of the crystal, as illustrated in Fig.~\ref{fig:CavityStructure}(a)~\cite{Woldering2014PRB}. 
Up to five cavity defect bands were found inside the band gap for defect pores smaller than all other pores, corresponding to donor states~\cite{footnote:AccepterStates}. 
The occurrence of five resonances suggests that the defect cavity has d-like character - in analogy to an atomic defect in a semiconductor band gap~\cite{Ashcroft1976Book} - or quadrupole character in terms of electrodynamic resonances~\cite{Jackson1999Book}, as is also borne out of a parallel theoretical study of a 3D superlattice of cavities~\cite{Hack2018}. 
Each of the resonances has a resonance frequency $\omega_m$~($m = 1...5$) that is nearly independent on wave vector, as expected for a defect state~\cite{Woldering2014PRB}. 
Since the calculations in Ref.~\cite{Woldering2014PRB} were performed for infinite crystals with no surrounding vacuum, however, the cavity quality factor could not be calculated and thus the energy enhancement and potential absorption in such a cavity could not be assessed. 

Therefore, we investigate here the optical properties of a 3D photonic band gap cavity with finite support that is surrounded by vacuum as in real devices, as guidance for experimental work. 
By cross-correlating the fields for a crystal with finite support with those of the defect states in the infinite crystal, we identify the resonances in the crystal with finite support and their field patterns. 
We verify the angle-independence of the reflectivity resonances to confirm the 3D localization of the cavity resonances in real space. 
We study the quality factors of the resonances, and calculate the electric-field energy enhancement due to these resonances and assess their potential application to enhance absorption, notably for photovoltaic applications. 
We also address resonances below the 3D band gap of the perfect crystal structure, and find evidence for Fano resonances.

\section{Methods}
The two 2D arrays of pores in a 3D inverse woodpile crystal structure have a radius $r$ and run in the orthogonal $X$ and $Z$ directions~\cite{Ho1994SSC}. 
The 2D arrays have a centered-rectangular lattice with lattice parameters $c$ (in the $X$ and $Z$-directions) and $a$ (in the $Y$-direction). 
The diamond-like structure is cubic when $\frac{a}{c} = \sqrt{2}$. 
For pores with a relative radius $\frac{r}{a} = 0.24$, cubic inverse woodpile crystals have a maximum band gap width with a broad relative bandwidth $\Delta \omega / \omega_{c} = 25.3\%$ relative to the central band gap frequency $\omega_{c}$~\cite{Hillebrand2003JAP,Woldering2008Nanotechnology}. 
 
Woldering \textit{et al.}~\cite{Woldering2014PRB} calculated the photonic band structure for a supercell of the 3D inverse woodpile photonic crystal with a cavity by employing a plane-wave expansion (PWE) method~\cite{Johnson2001OptExpress} that assumes the structure to be infinitely extended. 
They reported that defect pores with radius $r' = 0.5r$ yield an optimal light confinement. 
In order to relate to the previous work, we tuned the parameters to be the same as previously, namely an optimal ideal pore radius $\frac{r}{a} = 0.24$~\cite{Hillebrand2003JAP,Woldering2009JAP}, an optimal defect pore radius $\frac{r'}{r} = 0.50$ (or $\frac{r'}{a} = 0.12$)~\cite{Woldering2014PRB}, and a dielectric permittivity $\epsilon = 12.1$ that is typical for silicon in the near infrared and telecom ranges~\cite{Woldering2009JAP, Woldering2014PRB, Huisman2011PRB, Leistikow2011PRL}.

To investigate the consequences of the finite support and of the defect cavity, we study the reflectivity spectra for a 3D photonic band gap crystal with a point defect and with a finite thickness (slab geometry). 
Since our recent results revealed that a thin perfect crystal with a thickness of only three unit cells is sufficient to reveal strong reflectivity and strongly attenuated transmission~\cite{Devashish2017PRB}, we selected $3\times3\times3$ as the size of the super cell, as shown in Fig.~\ref{fig:CavityStructure}, to keep the computations tractable.
Thus, there is one point defect in the direction of propagation of the incident plane wave in the thickness $L_{Z} = 3c$ of the photonic crystal slab. 

We employ the finite-element method (FEM) to solve the time-harmonic Maxwell equations (using the commercial solver COMSOL~\cite{COMSOLMultiphysics}). 
In order to describe reflectivity (or transmission) from a photonic crystal slab oriented perpendicular to the $Z$ direction, we employ Bloch-Floquet periodic boundaries in the $\pm X$ and the $\pm Y$ directions and absorbing boundaries in the $-Z$ and $+Z$ directions~\cite{Joannopoulos2008Book}. 
The incident field starts from a plane in the $-Z$ direction that is separated from the crystal by an air layer. 
The plane represents a boundary condition rather than a true current source since it also absorbs the reflected waves~\cite{Jin1993Book}. 
We launch incident plane waves with either $s$ polarization - with the electric field perpendicular to the plane of incidence - or $p$ polarization - with magnetic field normal to the plane of incidence - and with an angle of incidence between $0\degree$ and $80\degree$. 

We investigate the effects of the symmetry-disruption of the infinite crystal by the interfaces by comparing the reflectivity spectra for a thin slab to the corresponding photonic band structure for an infinitely extended crystal~\cite{Devashish2017PRB}. 
To eliminate possible deviations arising from differences in numerical methods or in the detailed dielectric permittivity distributions $\epsilon(\vec{r})$, we employ the eigenvalue solver of our FEM solver~\cite{COMSOLMultiphysics} to compute the photonic band structure. 
Differences occurring between band structures computed with the plane-wave expansion (PWE) method and with FEM are discussed in Appendix~\ref{sect:PhotonicBandstructure}. 
Since a photonic band structure pertains to an infinitely extended crystal, we alter our finite reflectivity computational cell by employing Bloch-Floquet periodic boundaries in all three dimensions $\pm X$, $\pm Y$, and $\pm Z$~\cite{Joannopoulos2008Book}. 
Hence, the point defect sits at the center of a $3 \times 3 \times 3$ supercell that is replicated infinitely. 
  
We use tetrahedra as basic elements in our finite element mesh to subdivide the 3D computational cell into elements. 
To accurately mesh sharp interfaces in a 3D inverse woodpile crystal, an upper limit $\triangle l \leq \frac{\lambda_{0}}{8 \sqrt{\epsilon_{\textrm{max}}}}$ is imposed to the edge length $\triangle l$ on any tetrahedron in the inverse woodpile, with $\epsilon_{\textrm{max}}$ the maximum dielectric permittivity for the selected range of frequencies. 
An analysis of the mesh convergence is presented in Appendix~\ref{sect:MeshConvergence}. 
For comparison, the PWE method requires Cartesian meshing which results in a less faithful discretization of the crystal structure, and hence greater systematic errors than with the adaptive mesh in FEM, as discussed in Appendix~\ref{sect:PhotonicBandstructure}.

To detect narrow reflectivity resonance troughs (as analyzed in Appendix~\ref{sect:FrequencyConvergence}), we used a frequency resolution $\delta\tilde{\omega} = 0.001$~\cite{footnote:ReducedFrequency} below the 3D band gap and $\delta\tilde{\omega} = 0.0005$ in the 3D band gap, therefore we were able to resolve quality factors as high as $Q = 1000$. 
All calculations are performed on the ``Serendipity" cluster in the MACS group at the MESA$^{+}$ Institute~\cite{Serendipity}. 
Even on this powerful computer cluster, the computation of a single spectrum as shown in Fig.~\ref{fig:reflectivity_BandGap_S_P} takes more than $122$ hours or $6$ days.

\section{Results}

\subsection{Reflectivity resonances within the 3D band gap}
\label{subsect:WithinBandGap}
\begin{figure}[tbp!]
\centering
\includegraphics[width=1.0\columnwidth]{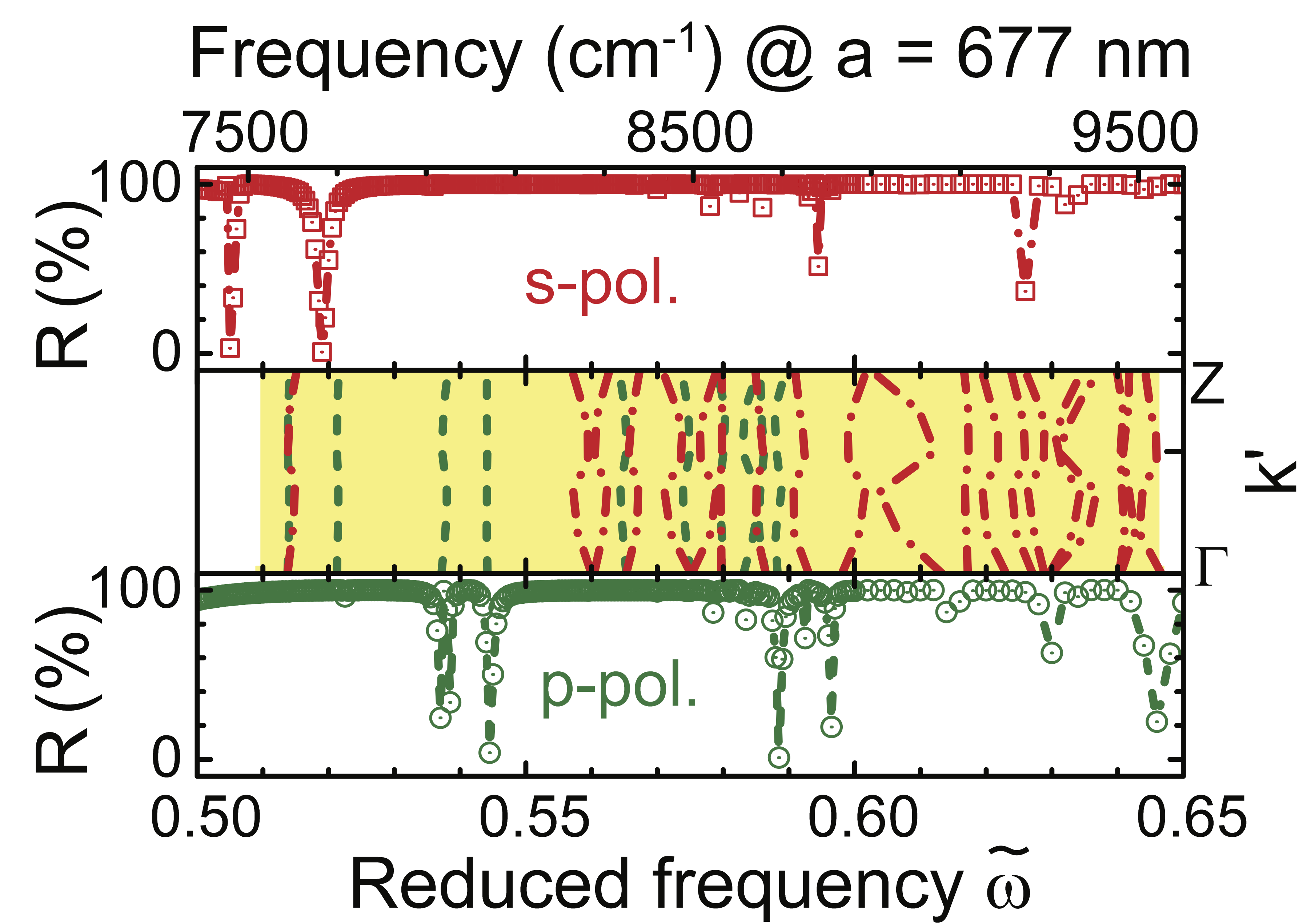}
\caption{Reflectivity spectra in the range of the 3D band gap calculated for a 3D inverse woodpile photonic crystal with a point defect (as shown in Fig.~\ref{fig:CavityStructure}(b)) for light at normal incidence ($\Gamma Z$ direction) for $s$ polarization (top panel, red dashed-dotted curve) and for $s$ and $p$ polarizations (bottom panel, green dashed curve). 
Central panel: band structure for wave vectors between $\Gamma$ and $Z$, where the 3D band gap of the perfect crystal is indicated with the yellow bar. 
Red dashed-dotted lines are the $s-$polarized bands and green dashed lines are $p-$polarized bands. 
The reduced frequency $\tilde{\omega}$ is defined in footnote~\cite{footnote:ReducedFrequency}. } 
\label{fig:reflectivity_BandGap_S_P}
\end{figure}
The central panel of Fig.~\ref{fig:reflectivity_BandGap_S_P} shows the polarization-resolved band structure in the $\Gamma Z$ high symmetry direction for a $3 \times 3 \times 3$ supercell of the 3D inverse woodpile photonic crystal with a point defect. 
For reference, the 3D photonic band gap of the perfect crystal spans from $\tilde{\omega} = 0.51$ to $\tilde{\omega} = 0.645$. 
With increasing frequency from the bottom of the band gap, we observe that there is one isolated $s$-polarized defect band S1 at $\tilde{\omega}_{1} = 0.5144$ and four isolated $p$-polarized defect bands P1, P2, P3, and P4 at $\tilde{\omega}_{2} = 0.5140,~\tilde{\omega}_{3} = 0.5213,~\tilde{\omega}_{4} = 0.5376,~\textrm{~and~}\tilde{\omega}_{5} = 0.5441$, respectively, where the S1 and P1 bands are nearly degenerate. 
The observation of five nearly dispersionless defect bands agrees very well with Ref.~\cite{Woldering2014PRB} who also reported five defect states, although the polarization of the states was not identified~\cite{footnote:ComparisonToHack2018}. 
The central panel of Fig.~\ref{fig:reflectivity_BandGap_S_P} reveals numerous other $s$ and $p-$defect bands beyond $\tilde{\omega} = 0.55$ that are possibly due to waveguiding along the defect pores. 
Since these bands are neither isolated nor dispersionless, we do not consider these to be cavity resonances, similar to Ref.~\cite{Woldering2014PRB}. 
For clarity, we note that these defect modes are fully vectorial in character. 
In this paper, we refer to them as $s$ or $p$-polarized only to indicate their symmetry properties, which allow them to be excited either with $s$ or $p$-polarized incident light. 
The frequencies reported in Ref.~\cite{Woldering2014PRB} are nearly $\triangle \tilde{\omega} = 0.007$ higher than the present results, which is attributed to the different numerical methods (PWE versus FEM) with concomitant different spatial meshing and resolutions, see Appendix~\ref{sect:PhotonicBandstructure}. 

Naively, one may expect a resonance trough in reflectivity for each of the five defect bands. 
To identify such resonance troughs, we calculate reflectivity inside the 3D band gap at normal incidence for a $3 \times 3 \times 3$ supercell of the 3D inverse woodpile photonic crystal with a point defect. 
The top and bottom panels of Fig.~\ref{fig:reflectivity_BandGap_S_P} show the reflectivity spectra for $s$ and $p$ polarizations, respectively. 
For both polarizations, however, we observe only four reflectivity resonances, namely one $s$-polarized one at $\tilde{\omega} = 0.519$, and three $p$-polarized ones at $\tilde{\omega} = 0.538,~0.540,~0.545$. 
Moreover, it is remarkable that there do not seem to be reflectivity counterparts for the S1-band at $\tilde{\omega}_{1} = 0.5144$, and for the two $p$-bands at $\tilde{\omega}_{2} = 0.5140$ and $\tilde{\omega}_{3} = 0.5213$. 
Thus, this straightforward inspection of the reflectivity spectra does not allow to identify reflectivity resonances in the finite-support crystal with defect states in its infinite counterpart, and a more advanced approach is needed. 

\begin{figure}[tbp!]
\centering
\includegraphics[width=1\columnwidth]{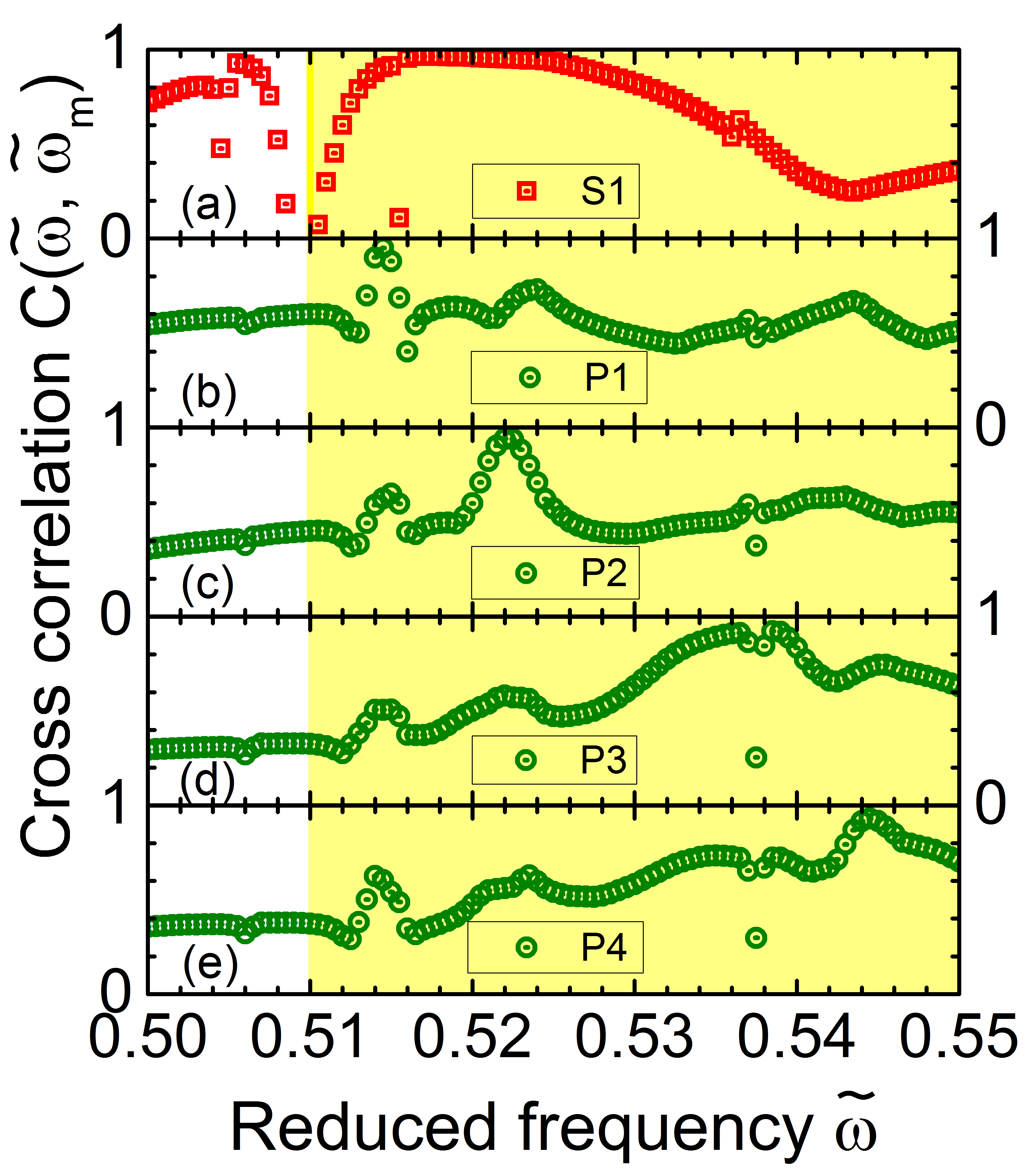}
\caption{
Normalized cross-correlation between the cross sections of the electric-field distribution ($|{E}|$) at the defect bands $\tilde{\omega}_{m}$ in the band structure and fields at discrete frequencies $\tilde{\omega}$ in the reflectivity spectra for a finite 3D inverse woodpile photonic crystal with a point defect. 
The 3D band gap of the perfect crystal is shown with the yellow bar. 
S1 indicates the single $s-$polarized cavity resonance inside the 3D band gap. 
Similarly, P1, P2, P3, and P4 indicate the four $p-$polarized cavity resonances inside the 3D band gap. 
} 
\label{fig:EfieldCorrelation}
\end{figure}

To match a reflectivity trough of a finite crystal slab to a corresponding defect band of an infinite crystal, we cross-correlate the spatial distribution of the electric-field norm $|{E}(\tilde{\omega}_{m}, \vec{r})|$ of a defect band at frequency $\tilde{\omega}_{m}$ with the field $|{E}(\tilde{\omega}, \vec{r})|$ in the finite crystal slab calculated as a function of frequency $\tilde{\omega}$. 
To keep the computation tractable~\cite{footnote:NormEField}, we consider the field distributions  $|{E}(\tilde{\omega}_{m}, y, z, \vec{r_0})|$ and $|{E}(\tilde{\omega}, y, z, \vec{r_0} + \vec{\Delta r})|$ in the $YZ$ cross section through the center of the cavity at the reference position $\vec{r_0}$ for the infinite crystal and at $(\vec{r_0} + \vec{\Delta r})$ for the finite-support crystal, see Fig.~\ref{fig:normE_Band_Trough_P} for such field patterns. 
The normalized cross-correlation $C(\tilde{\omega}, \tilde{\omega}_{m}, \vec{\Delta r}))$ is defined as

\begin{align}
\begin{split}
& C(\tilde{\omega}, \tilde{\omega}_{m}, \vec{\Delta r}) \equiv\\
&\int_{z = 0}^{3c} \int_{y = 0}^{3a} \Big( |{E}(\tilde{\omega}_{m}, y, z, \vec{r_0})| |{E}(\tilde{\omega}, y, z, \vec{r_0} + \vec{\Delta r})| \Big) dy dz \\
& \cdot \frac{1}{\Big[\int_{z = 0}^{3c} \int_{y = 0}^{3a} (|{E}(\tilde{\omega}_{m}, y, z, \vec{r_0})|)^2 dy dz\Big]^{\frac{1}{2}}} \\
& \cdot \frac{1}{\int_{z = 0}^{3c} \int_{y = 0}^{3a} (|{E}(y, z, \vec{r_0} + \vec{\Delta r}, \tilde{\omega})|)^2 dy dz \Big]^{\frac{1}{2}}},
\label{eq:C_full}
\end{split}
\end{align}
 
When the position of the cavity in the finite-support crystal is matched to the position in the infinite crystal, $\vec{\Delta r} = 0$, and $C(\tilde{\omega}, \tilde{\omega}_{m}, \vec{\Delta r})$ is at a maximum; in Fig.~\ref{fig:EfieldCorrelation} we plot $C(\tilde{\omega}, \tilde{\omega}_{m}, \vec{\Delta r} = 0)$. 
A reflectivity resonance with a field that cross-correlates to $C = 1$ at its central frequency $\tilde{\omega}_{c}$ will correspond to a defect state at $\tilde{\omega}_{m}$. 
In addition to the field cross-correlation, we also confirm the cavity resonance by the visual inspection of the cross sections of the electric-field distributions, and we verify the 3D band gap confinement of light by checking the angular independence of a reflectivity resonance. 
From the five resonances S1, P1, P2, P3, and P4, it appears that P1, P2, and P4 show relatively straightforward behavior, therefore we first discuss these three resonances, before analyzing the more complex behavior of S1 and P3 resonances. 

\begin{figure}[tbp!]
\centering
\includegraphics[width=1\columnwidth]{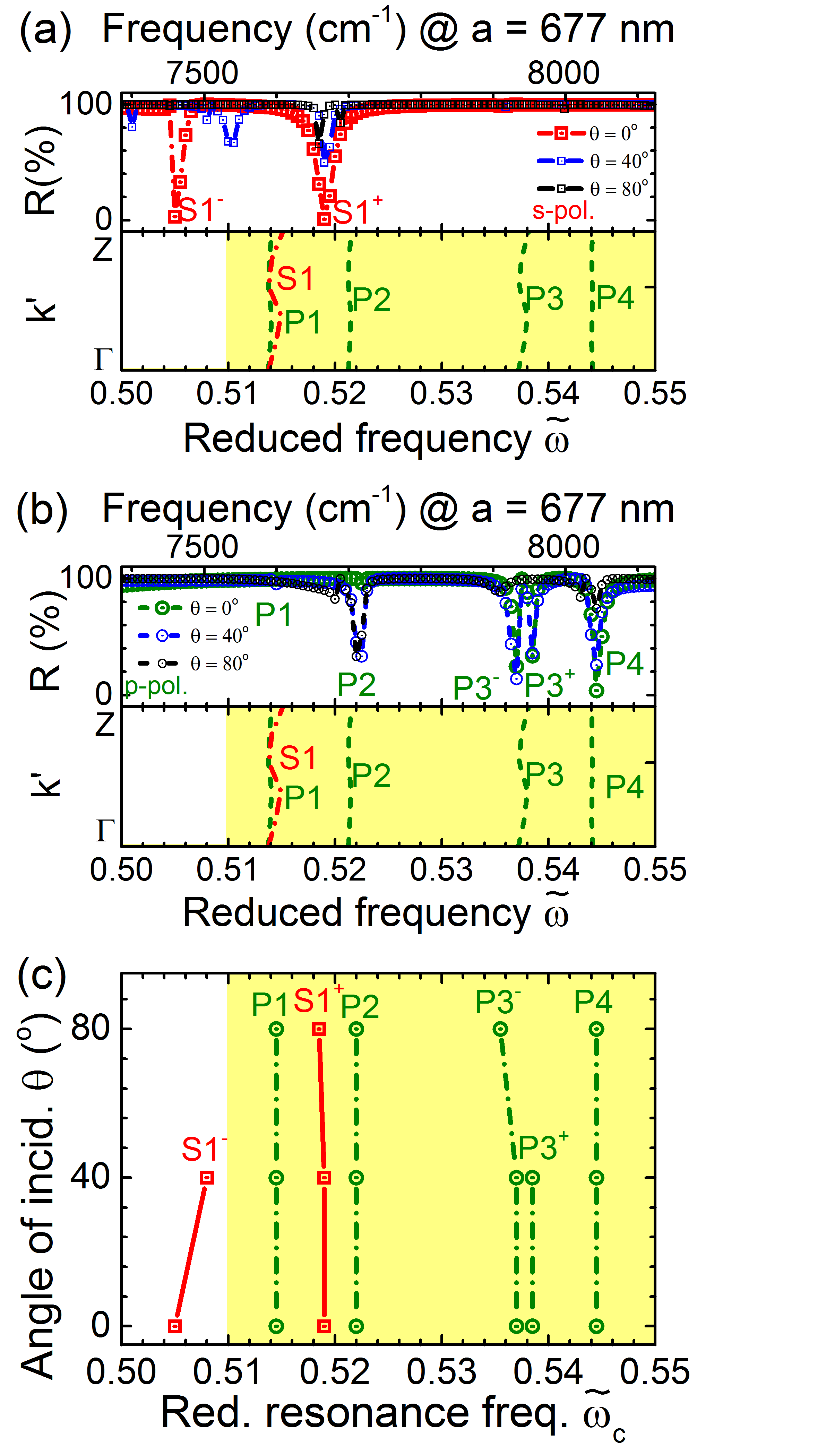}
\caption{Reflectivity near the bottom of the 3D band gap for (a) $s-$ and (b) $p-$polarized light for a 3D inverse woodpile photonic crystal with a point defect. 
In both upper panels, the red dashed-dotted and green dashed curves are calculated results at normal incidence, as in Fig. \ref{fig:reflectivity_BandGap_S_P}. 
Blue solid and black curves in the upper panels are reflectivity spectra for angles of incidence $40 \degree$ and $80 \degree$, respectively. 
Both lower panels show the band structures in the $\Gamma Z$ direction for (a) $s-$ (red) and (b) $p-$polarized light (green). 
The wave vector is reduced as $k^{'} = (ka/2\pi)$. 
The 3D band gap of a perfect infinite crystal is indicated with the yellow bar. 
(c) The central frequency of a resonance trough versus angle of incidence. 
S1$^{-}$, S1$^{+}$, P1, P2, P3$^{-}$, P3$^{+}$, and P4 label reflectivity resonances. 
}  
\label{fig:ReflectivityAngle_S_P}
\end{figure}

\begin{figure}[tbp!]
\centering
\includegraphics[width=1\columnwidth]{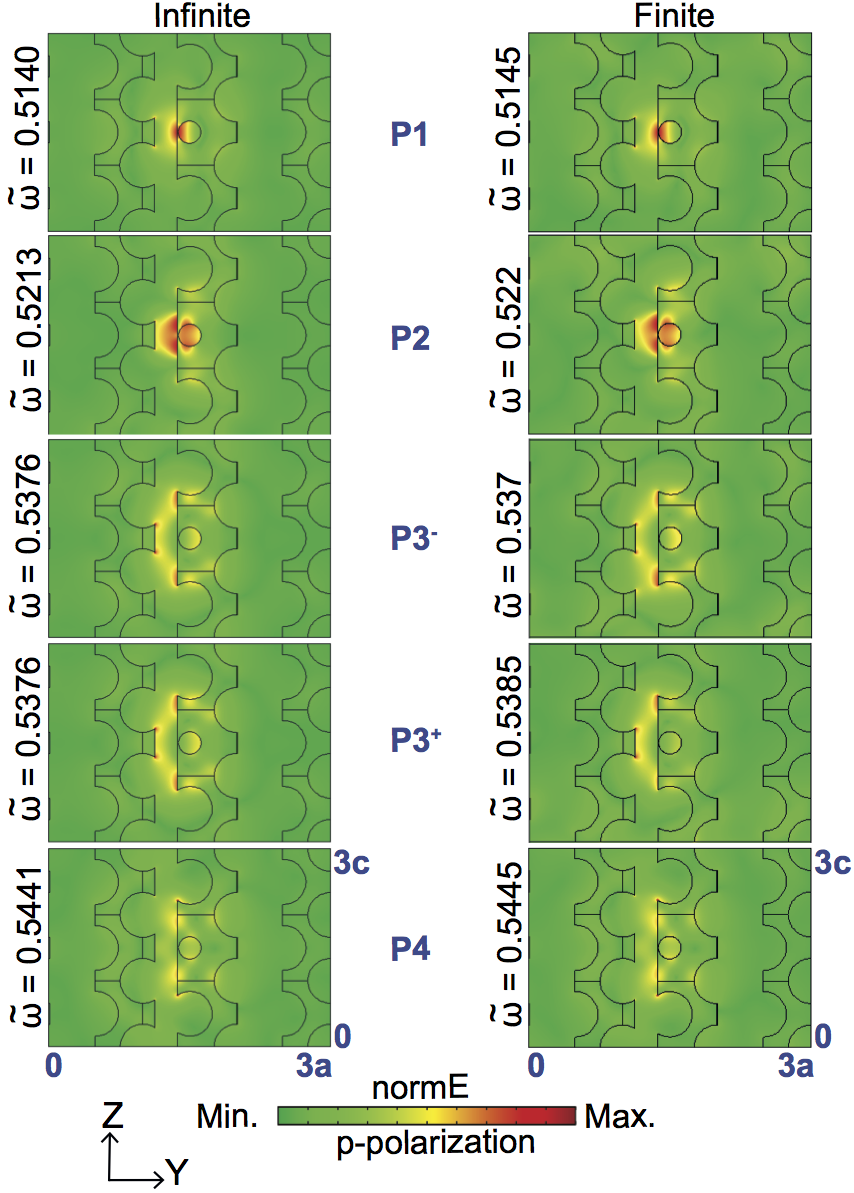}
\caption{
Left: $YZ$ cross sections of the spatial distribution of the electric-field modulus ($|E|$) of the P1, P2, P3, and P4 resonances in an infinitely replicated $3\times3\times3$ super cell of a 3D inverse woodpile photonic crystal with a point defect. 
Right: $YZ$ cross sections of the electric-field modulus ($|E|$) for the P1, P2, P3$^{-}$, P3$^{+}$, and P4 resonances in a 3D inverse woodpile photonic crystal with a point defect with finite support for $p-$polarized light at normal incidence and a $3\times3\times3$ super cell. 
} 
\label{fig:normE_Band_Trough_P}
\end{figure}
\begin{figure}[tbp!]
\centering
\includegraphics[width=1\columnwidth]{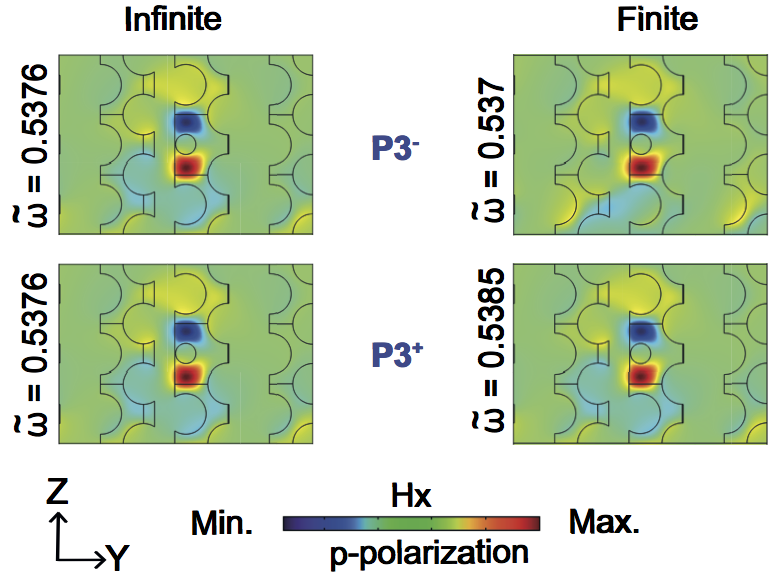}
\caption{
Left: $YZ$ cross section of the spatial distribution of the x-component of the magnetic-field ($Hx$) for the P3 defect state in an infinitely extended $3\times3\times3$ super cell of a 3D inverse woodpile photonic crystal with a point defect. 
Right: $YZ$ cross section of the magnetic-field ($Hx$) for the P3$^{-}$ and P3$^{+}$ resonances in a 3D inverse woodpile photonic crystal with a point defect with finite support for $p-$polarized light at normal incidence and a $3\times3\times3$ super cell. 
} 
\label{fig:normH_Band_Trough_P3}
\end{figure}

\textbf{P1 resonance:} 
In Fig.~\ref{fig:EfieldCorrelation}(b), the cross-correlation of the field of the P1 defect band at $\tilde{\omega}_{m = 2} = 0.5140$ with the finite-crystal fields equals about 0.6 at most frequencies, with a marked peak at $\tilde{\omega} = 0.5140$. 
This cross-correlation peak with $C = 1$ corresponds to a weak resonance trough in reflectivity ($R_{min} \simeq 99\%$) at normal incidence. 
Upon varying the angle of incidence, we observe that the reflectivity resonance becomes a bit more prominent ($R_{min} \simeq 95\%$), and that the reflectivity resonance frequency is independent of angle, as shown in Fig.~\ref{fig:ReflectivityAngle_S_P}. 
In addition, Fig.~\ref{fig:normE_Band_Trough_P} shows that the spatial $YZ$ field distribution for the defect band in the infinite crystal matches very well with the field distribution of the resonance in the photonic crystal with finite support. 
Therefore, from all three main observations (correlation, angle-independent reflectivity resonance, and spatial field distribution) we identify the P1 defect band to occur in reflectivity at $\tilde{\omega}_{m} = 0.5140$ with a field distribution shown in Fig.~\ref{fig:normE_Band_Trough_P}. 

\textbf{P2 resonance:} 
The cross correlation of the field of the P2 defect band at $\tilde{\omega}_{m = 3} = 0.5213$ in the infinite crystal is nearly constant at about $C = 0.4$, with a maximum $C = 1$ at a frequency $\tilde{\omega} = 0.522$ that agrees well with a weak reflectivity resonance at $\tilde{\omega} = 0.522$ at normal incidence, see Fig.~\ref{fig:ReflectivityAngle_S_P}(b). 
With increasing angle of incidence the reflectivity resonance deepens to $R_{min} \simeq 30\%$. 
The frequency of the reflectivity resonance is independent of incident angle (Figure~\ref{fig:ReflectivityAngle_S_P}) as expected for a cavity resonance in a 3D band gap. 
Figure~\ref{fig:normE_Band_Trough_P} shows that the field distribution in the infinite crystal agrees very well with the field distribution in the finite crystal slab. 
We note that the P2 field distribution resembles the P1 field distribution, since there is a secondary cross-correlation peak at the P1-resonance at $\tilde{\omega} = 0.514$ in Fig.~\ref{fig:EfieldCorrelation}. 
Thus, we identify the P2 resonance to occur at $\tilde{\omega}_{m} = 0.522$ with a field distribution as shown in Fig.~\ref{fig:normE_Band_Trough_P}. 

\textbf{P4 resonance:} 
The cross correlation of the P4 defect band at $\tilde{\omega}_{m = 5} = 0.5441$ in the infinite crystal (Figure~\ref{fig:reflectivity_BandGap_S_P} with the finite crystal field distributions is nearly constant at about $C = 0.3$ (Figure~\ref{fig:EfieldCorrelation} (e)). 
At $\tilde{\omega} = 0.5445$, the cross correlation has a maximum at $C = 1$ that agrees well with a strong reflectivity ($R_{min} \simeq 0\%$) resonance at $\tilde{\omega} = 0.5445$ at normal incidence (see Fig.~\ref{fig:ReflectivityAngle_S_P}(b)). 
With increasing angle of incidence the reflectivity resonance changes to $R_{min} \simeq 70\%,$ or $20\%$. 
The reflectivity resonance is independent of incident angle (Fig.~\ref{fig:ReflectivityAngle_S_P}), as expected for a cavity resonance in a 3D band gap. 
Finally, Fig.~\ref{fig:normE_Band_Trough_P} shows that the spatial field distribution in the infinite crystal defect band agrees very well with the field distribution in the finite crystal slab. 
The cross-correlation shows that the P4 field distribution resembles the P3 field distribution in view of a secondary cross-correlation peak at $\tilde{\omega} = 0.537$ and $\tilde{\omega} = 0.5385$ in Fig.~\ref{fig:EfieldCorrelation}. 
Thus, we identify the P4 defect band to occur at $\tilde{\omega}_{m} = 0.5445$ with a field distribution as shown in Fig.~\ref{fig:normE_Band_Trough_P}. 

\textbf{P3 resonance:} 
The P3 defect band at $\tilde{\omega}_{m = 4} = 0.5376$ in the infinite crystal reveals a surprising double cross correlation peak with $C = 1$ at $\tilde{\omega}_{-} = 0.537$ and $\tilde{\omega}_{+} = 0.5385$, see Fig.~\ref{fig:EfieldCorrelation}. 
Indeed the reflectivity spectrum also shows two corresponding resonance troughs at $\tilde{\omega}_{-} = 0.537$ and $\tilde{\omega}_{+} = 0.5385$, see Fig.~\ref{fig:ReflectivityAngle_S_P}(b). 
These two troughs are symmetrically located on either side of the defect band $\tilde{\omega_{m}} = 0.537$. 
Both troughs develop deep minima ($R_{min} \leq 20 \%$) versus angle of incidence, and their resonance frequencies are independent of angle (Fig.~\ref{fig:ReflectivityAngle_S_P}), typical of 3D photonic band gap cavity resonances. 
Figure~\ref{fig:normE_Band_Trough_P} shows that the spatial electric-field distribution of the P3 defect band matches very well with both field distributions at the lower ($\tilde{\omega}_{-} = 0.537$) and upper ($\tilde{\omega}_{+} = 0.5385$) reflectivity resonances. 
Since the magnetic field of an electromagnetic wave is complementary to the electric field and thus holds additional information~\cite{Griffiths1998Book, Burresi2009Science}, we show the magnetic-field distributions for both the infinite crystal and the finite-support crystal in Fig.~\ref{fig:normH_Band_Trough_P3}. We see that both the field distributions and the phases of both finite-crystal resonances match very well with the distribution and phase of the single P3 defect state in the infinite crystal, which confirms that both finite-size resonances derive from one and the same resonance in the infinite crystal. 
\begin{figure}[tbp!]
\centering
\includegraphics[width=1\columnwidth]{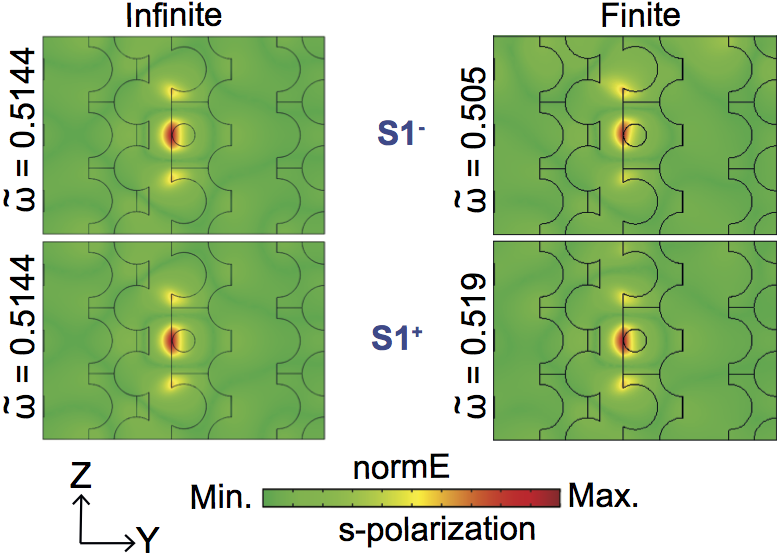}
\caption{
Left: $YZ$ cross section of the spatial distribution of the electric-field modulus ($|E|$) in an infinitely extended $3\times3\times3$ super cell of a 3D inverse woodpile photonic crystal with a point defect. 
Right: $YZ$ cross sections of the same electric-field moduli for the S1$^{-}$ and S1$^{+}$ resonances in a 3D inverse woodpile photonic crystal with a point defect with finite support for $s-$polarized light at normal incidence and a $3\times3\times3$ super cell. 
} 
\label{fig:normE_Band_Trough_S}
\end{figure}

\textbf{S1 resonance:} 
The S1 field distribution at $\tilde{\omega}_{m = 1} = 0.5144$ in the infinite crystal also reveals a double cross correlation peak in Fig.~\ref{fig:EfieldCorrelation} with maxima $C = 1$ at $\tilde{\omega}_{-}  = 0.505$ and at $\tilde{\omega}_{+} = 0.5190$, the latter being remarkably broad. 
In reflectivity (Fig.~\ref{fig:reflectivity_BandGap_S_P}(a)) we observe two matching resonances at normal incidence at $\tilde{\omega}_{-}  = 0.505$ and $\tilde{\omega}_{+}  = 0.5190$. 
With increasing angle of incidence the two reflectivity resonances reveal remarkably different behavior: the upper resonance at $\tilde{\omega}_{+} = 0.5190$ does not shift with angle of incidence (Fig.~\ref{fig:ReflectivityAngle_S_P}), as expected for a photonic band gap cavity. 
The lower resonance at $\tilde{\omega}_{-}  = 0.505$ (at $\theta = 0\degree$), however, shifts with angle of incidence since different resonances occur at $\theta = 40\degree$ or $\theta = 80\degree$. 
The shift makes intuitive sense in view of the striking occurrence of this resonance \textit{outside} the photonic band gap, hence no 3D confinement is expected. 
Nevertheless, this resonance occurs inside the $s-$polarized $\Gamma Z$ stop band so at least one-dimensional (1D) confinement is expected. 
Another remarkable feature of the S1 resonance is that the two reflectivity resonances are asymmetrically located on either sides of the defect state in the infinite-crystal at $\tilde{\omega_{m}} = 0.5144$. 

At this time, we do not yet have a physical explanation for the remarkable frequency splitting of the S1 and P3 defect states in the crystal with finite support. 
We speculate whether a hybridization with surface modes or the leaking of modes of the surrounding vacuum into the finite crystal slab may cause the splitting, although we are unaware why or how there could be resonant structure in the vacuum. 
It is also conceivable that the double reflectivity resonances are the result of the coupling of the defect state with another, as yet unidentified, resonance. 

\subsection{Quality factor and energy enhancement}
\label{subsect:FieldEnhancement}

\textbf{Quality factor:} 
Since we consider here a 3D photonic crystal with finite support, our study allows us to investigate the quality factor $Q$ of all resonances. 
Since we perform steady-state computations, 
we calculate the cavity quality factor $Q$ from the ratio of the central frequency $\tilde{\omega}_{c}$ and the full width at half maximum of the reflectivity resonance $\Delta \tilde{\omega}$~\cite{Feynman1964Book, Griffiths1998Book}
\begin{align}
\textrm{Q} = \frac{\tilde{\omega}_{c}}{\Delta \tilde{\omega}},
\label{eq:QualityFactor}
\end{align}

\begin{figure}[tbp!]
\centering
\includegraphics[width=1.0\columnwidth]{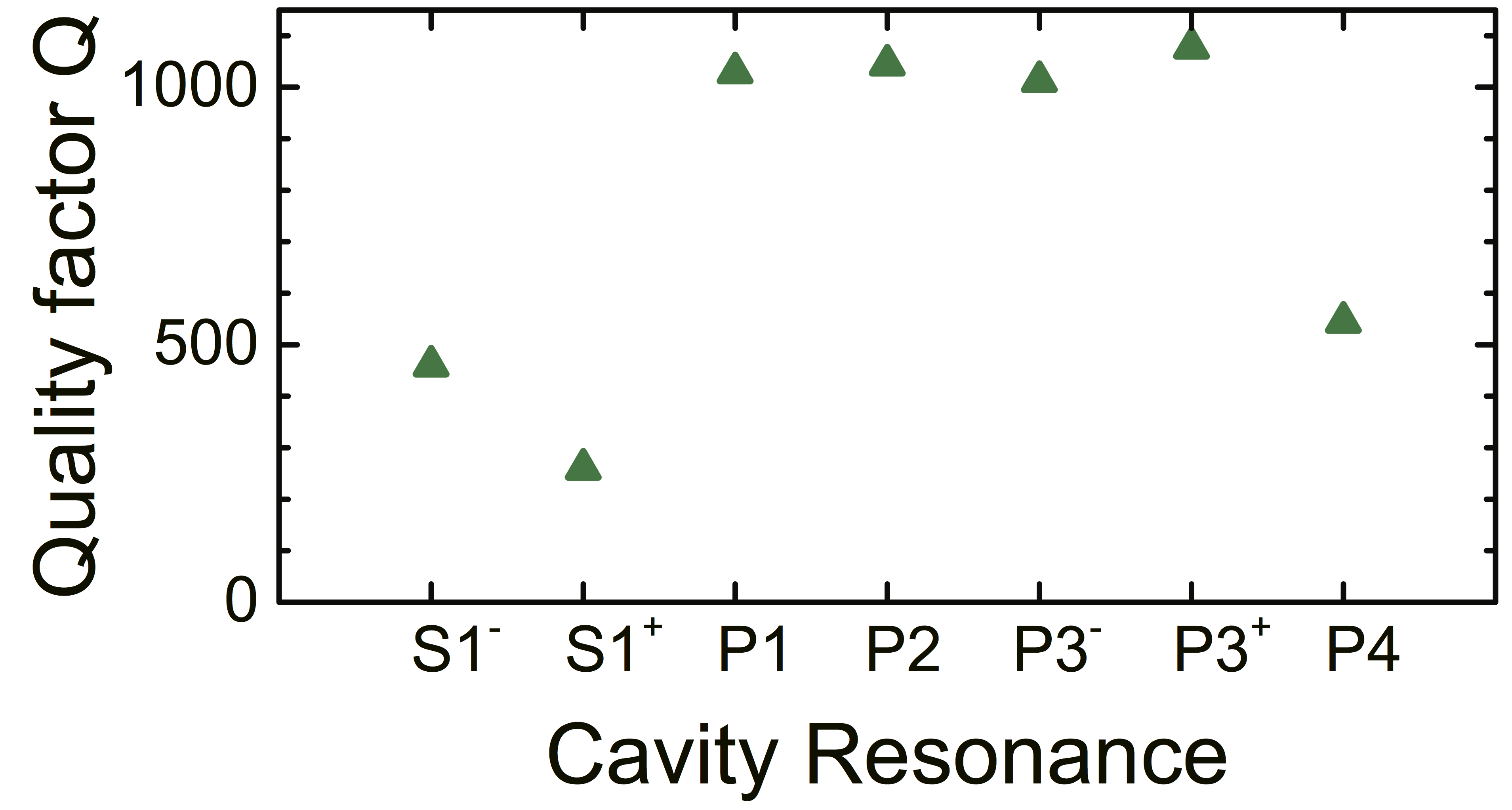}
\caption{Quality factor $Q$ (green triangles) for the reflectivity resonances S1$^{-}$, S1$^{+}$, P3$^{-}$, P3$^{+}$, and P4. 
The frequency resolution in the calculation places a lower bound Q$~=~1000$ to the quality factors of the P1 and the P2 resonances. 
} 
\label{fig:QMVFP}
\end{figure}
From the reflectivity spectra in Fig.~\ref{fig:ReflectivityAngle_S_P} we obtain the bandwidths 
of the resonances S1$^{-}$, S1$^{+}$, P3$^{-}$, P3$^{+}$~\cite{footnote:P3Bandwidth}, and P4 to be $\Delta \tilde{\omega} = 1.1 \times 10^{-3}, 2.0 \times 10^{-3}, 5.3 \times 10^{-3}, 0.5 \times 10^{-3}$, and $1.0 \times 10^{-3}$, respectively, which corresponds to the cavity quality factors $Q$ shown in Fig.~\ref{fig:QMVFP}. 
The P1 , P2,  P3$^{-}$, P3$^{+}$ resonances have the highest cavity quality factors near $Q = 1000$ and the S1$^{+}$ resonance has the lowest quality factor of about $Q = 250$. 
Since the bandwidth of the P1 and the P2 resonances is close to the numerical resolution, we take the minimum bound for their bandwidth to be the frequency resolution $\triangle \tilde{\omega} = \delta \tilde{\omega} = 0.0005$, and thus a corresponding quality factor $Q = 1000$ as a lower bound. 

For a strong confinement of light by an optical cavity, a large quality factor is most desirable~\cite{Vahala2003Nature,Gerard1998PRL}. 
Hence, the P2 reflectivity resonance with $\tilde{\omega_{c}} = 0.522$ has the best potential for the 3D spatial confinement of light. 
This conclusion matches with Ref.~\cite{Woldering2014PRB} though interestingly for a different underlying motivation: 
Woldering \textit{et al.} concluded that the $j=3$ resonance is the best, since it is most isolated as it has the greatest frequency difference to the other defect states~\cite{Woldering2014PRB}.\\

\begin{figure}[tbp!]
\centering
\includegraphics[width=1\columnwidth]{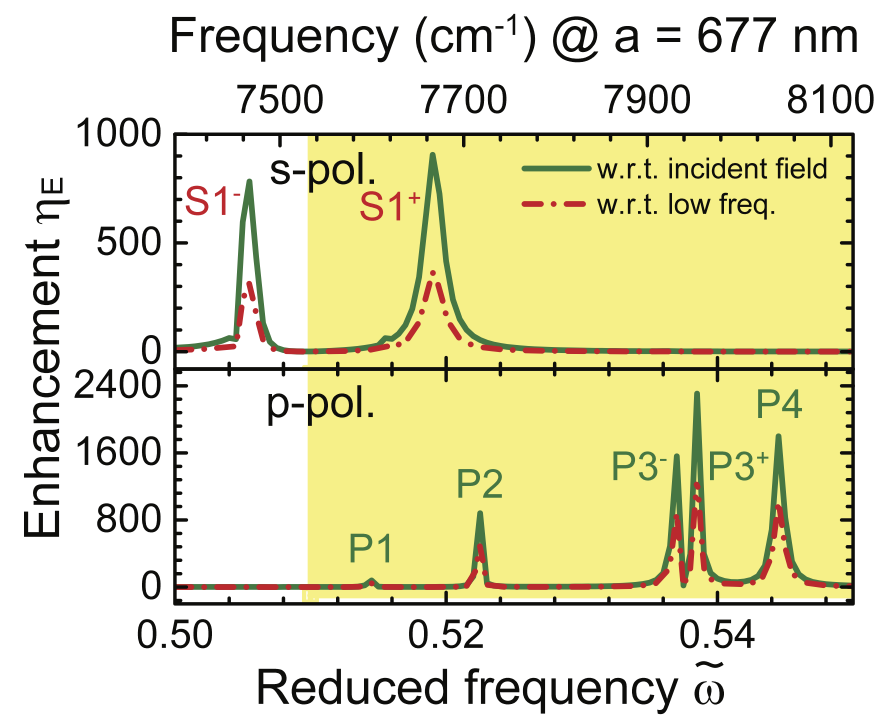}
\caption{Energy enhancement for a 3D inverse woodpile photonic crystal with a point defect. 
Red dashed-dotted curves represent the enhancement with respect to the effective medium approximation of the crystal taken at frequencies below the  band gap~\cite{Datta1993PRB}. 
Green solid curves represent the enhancement with respect to the incident light in vacuum. 
S1$^{-}$, S1$^{+}$, P3$^{-}$, P3$^{+}$, and P4 are reflectivity resonances identified above. 
} 
\label{fig:EnergyDensityEnhancement}
\end{figure}

\textbf{Energy enhancement:} 
Considering the electric-field distributions of Figs.~\ref{fig:normE_Band_Trough_P} and~\ref{fig:normE_Band_Trough_S}, we note that the electric field is strongly concentrated in the proximal region of the two orthogonal defect pores for all cavity resonances. 
We observe that the cavity resonances are localized both in silicon and in air, \textit{e.g.}, the S1$^{+}$ and P1 resonances have field maxima in the silicon backbone, whereas the P2 resonance has maxima in both air and in silicon. 
Therefore, to accurately quantify the energy enhancement $\eta_{E}$ at a frequency $\omega$ in the reflectivity spectra with respect to the reference frequency $\omega_{\textrm{ref}}$, we employ the definition
\begin{equation}
\eta_{E} \equiv \frac{\int_V \epsilon(\vec{r}) |E(\vec{r}, \omega)|^2 dV}{\int_V \epsilon(\vec{r}) |E(\vec{r}, \omega_{\textrm{ref}})|^2 dV}.
\label{eq:EnergyEnhancement}
\end{equation}
where we choose the integration volume as $V = ac^2$, the volume of one unit cell of the cubic inverse woodpile photonic crystal. 
To normalize the energy density, we consider two reference states, namely vacuum (by computing the energy density outside the crystal), or the homogeneous effective medium (by considering light at a low frequency $\tilde{\omega} = 0.04$ below the gap).
Figure~\ref{fig:EnergyDensityEnhancement} shows the enhancement $\eta_{E}$ between $\tilde{\omega} = 0.5$ and $\tilde{\omega} = 0.55$ for both polarizations. 
We observe large $\eta_{E} > 800$-fold energy enhancements at frequencies pertaining to the reflectivity resonances, \textit{i.e.}, S1$^{-}$, S1$^{+}$, P3$^{-}$, P3$^{+}$, and P4, which confirms our results regarding the identification of these resonances.
The enhancement of the S1$^{+}$ resonance is nearly equal to the one for the P2 resonance. 
The energy enhancement is maximum for the P3$^{-}$, P3$^{+}$, and P4 resonances, reaching up to $\eta_{E} = 2400 \times$.  

\subsection{Enhanced absorption}
\label{subsect:CavityPhotovoltaics}
To benefit from the large energy enhancement at the reflectivity resonances, we investigate the possibility of using a 3D silicon inverse woodpile with a resonant cavity as an absorbing medium in the visible part of the spectrum, for instance, for an optical sensor or a solar cell. 
It is well known that a thin submicron silicon film absorbs weakly in the wavelength range from $600$ to $1000$ nm~\cite{Chopra2004PPRA,Green2007JMS}.  
To enhance the absorption in this range we tailor the lattice parameters of the inverse woodpile to $a = 425$ nm and $c = 300$ nm such that the reflectivity resonances occur in this range. 
To make our calculations relevant to future experimental work, we employ a realistic refractive index of silicon including dispersion and absorption taken from Ref.~\cite{Green2008SolarEnergyMater}. 

\begin{figure}[tbp!]
\centering
\includegraphics[width=1\columnwidth]{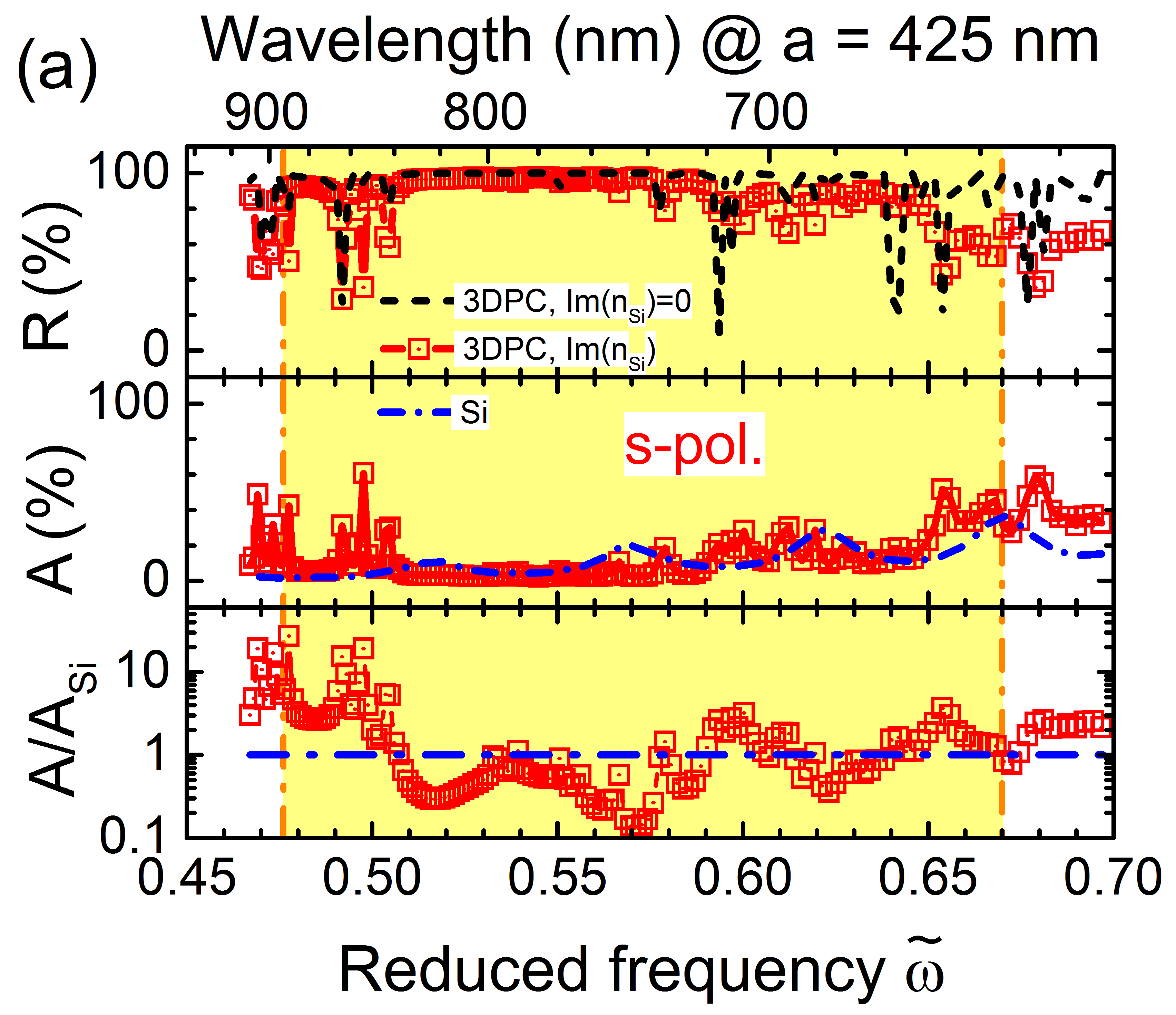}\\
\vspace{0.25cm}
\includegraphics[width=1\columnwidth]{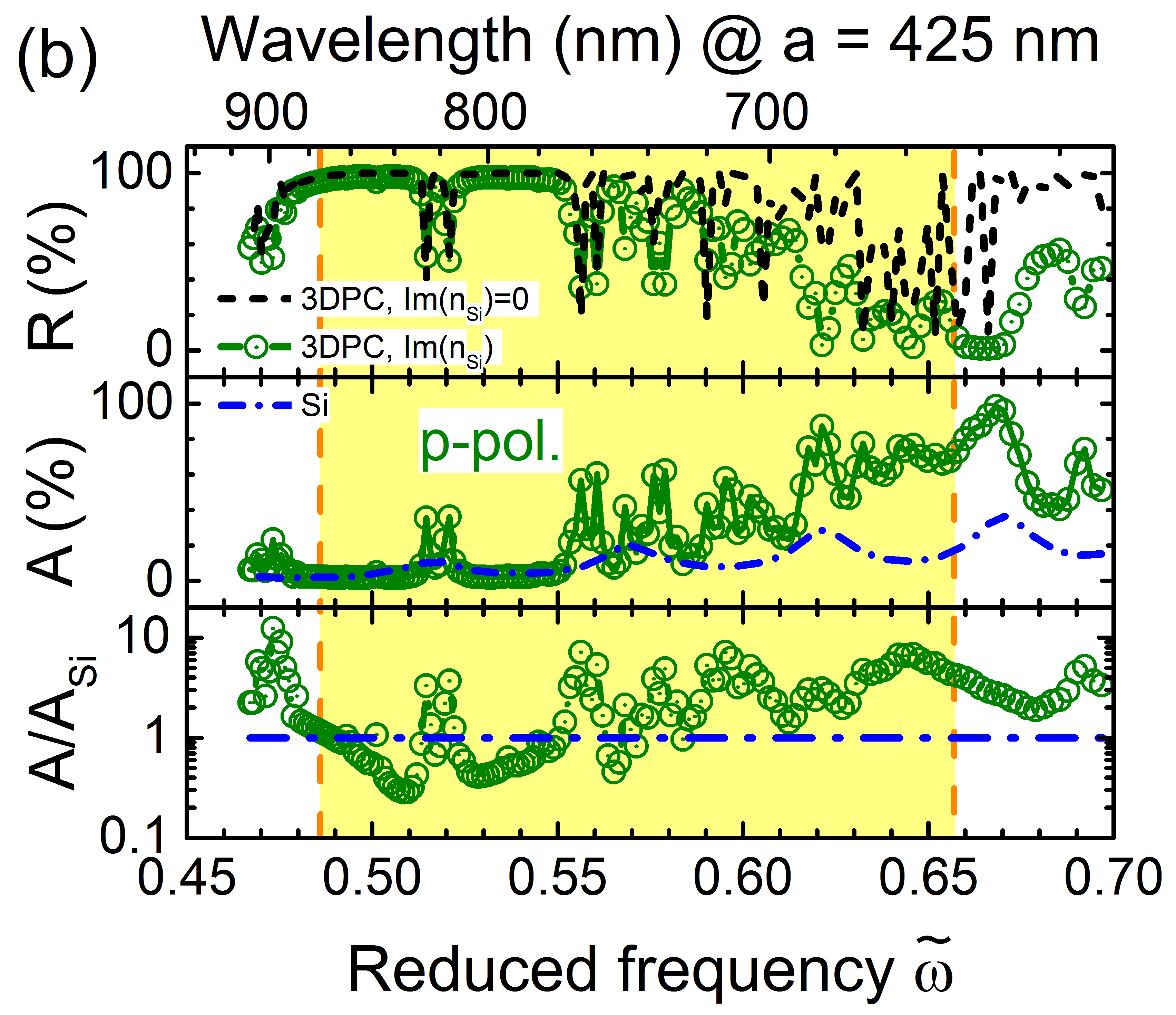}
\caption{Reflectivity, absorption, and absorption enhancement spectra calculated for a 3D inverse woodpile photonic crystal with proximal line defects (cavity) in the near IR and visible ranges at normal incidence in the $\Gamma Z$ direction for (a) $s$ polarization and (b) $p$ polarization. 
The yellow bar indicates the stop band for an inverse woodpile crystal without defects yet otherwise the same properties. 
Upper panels: the black dashed curves are reflectivity in case of dispersion but no absorption ($\mathbb{I}m(n_{Si}) = 0$). 
The red dashed-dotted and green dashed curves are reflectivity for both dispersion and absorption ($\mathbb{I}m(n_{Si}) \neq 0$). 
Middle panels: the red dashed-dotted and green dashed curves are absorption in case of both dispersion and absorption ($\mathbb{I}m(n_{Si}) \neq 0$). 
The blue dashed-dotted curves are absorption $A_{Si}$ for a thin homogeneous silicon film of $L_{Si} = 3$c $= 900$ nm. 
Bottom panels: the red dashed-dotted and green dashed curves are absorption enhancement of the crystal relative to the thin silicon film, the blue dashed-dotted lines indicate the reference level (= 1).  
} 
\label{fig:AbsorptionCavityResonance}
\end{figure}

Figures~\ref{fig:AbsorptionCavityResonance}(a,b) show reflectivity and absorption spectra for $s$ and $p$ polarizations, respectively. 
For an ideal inverse woodpile crystal with the same dispersive and complex refractive index, we find in separate calculations that the $s$-stop band appears between $\tilde{\omega} = 0.476$ and $\tilde{\omega} = 0.670$ (Fig.~\ref{fig:AbsorptionCavityResonance}(a)), and the $p$-stop band in between $\tilde{\omega} = 0.486$ and $\tilde{\omega} = 0.657$ (Fig.~\ref{fig:AbsorptionCavityResonance}(b)). 
Compared to an ideal inverse woodpile crystal with a refractive index typical of silicon in the near-infrared below its electronic bandgap including the telecom ranges (see Ref.~\cite{Devashish2017PRB}, Fig.~7), we find that the $s-$ and $p-$ stop bands in Fig.~\ref{fig:AbsorptionCavityResonance}(a,b) are shifted to lower reduced frequency. 
This shift makes sense since the refractive index in the visible range is greater than at optical frequencies below the electronic band gap~\cite{Green2008SolarEnergyMater}. 

For the inverse woodpile crystal with cavity, we observe in Fig.~\ref{fig:AbsorptionCavityResonance}(a,b) that there are numerous narrow resonance troughs in the absence of imaginary refractive index. 
For instance, for $s$ polarization we see narrow resonances at $\tilde{\omega} = 0.47$, $0.49$, $0.59$, $0.64$, and for $p$ polarization at $\tilde{\omega} = 0.47$, $0.52$, $0.55$, $0.64$. 
When we introduce the imaginary part of the silicon refractive index, there are still resonances that even appear at the same frequencies as for zero imaginary refractive index. 
To profit from the identification of the finite-support resonances in Section~\ref{subsect:WithinBandGap} where a purely real refractive index was used, we compare the reflectivity spectra in the top panels of Figs.~\ref{fig:AbsorptionCavityResonance}(a,b) both without and with imaginary part of the silicon refractive index. 
It is remarkable that in presence of the imaginary part of the refractive index several resonances are more pronounced than in absence of imaginary index, for instance, at $\tilde{\omega} = 0.50$ and $0.51$ for $s$ polarization and at $\tilde{\omega} = 0.57$ and $0.58$ for $p$ polarization. 
In terms of a 1D planar Fabry-Perot microcavity model~\cite{Gerard2003TAP}, we propose that the imaginary index effectively increases the reflectivity of the (photonic crystal) mirrors that surround the defect cavity, hence the resonance's amplitude increases. 

As a first step to investigate possible photovoltaic functionality of the 3D inverse woodpile photonic crystal, we calculate absorption and enhancement spectra that are shown in the middle and bottom panels of Figs.~\ref{fig:AbsorptionCavityResonance}(a), (b) for both $s$ and $p$ polarizations. 
We compare the absorption of an inverse woodpile crystal with thickness $L_{3DPC} = 3c = 900$ nm to the absorption of a thin homogeneous silicon film of equal thickness $L_{Si} = 900$ nm. 
We observe that at reduced frequencies beyond 0.55, an inverse woodpile crystal has substantially greater absorption than a homogeneous thin film, especially for $p$ polarization with enhancement as high as $10 \times$, see Figs.~\ref{fig:AbsorptionCavityResonance}(a), (b) bottom panels. 
The absorption peaks near $\tilde{\omega} = 0.49$ for $s$ polarization and $\tilde{\omega} = 0.52$ for $p$ polarization are similar to corresponding cavity resonances identified above in Section~\ref{subsect:WithinBandGap}.  

In addition to remarkable absorption, a 3D inverse woodpile photonic crystal has the advantageous feature that it is much rarified compared to bulk silicon, as it contains only $20 \%$ volume fraction silicon~\cite{Devashish2017Thesis} and is thus $4 \times$ lighter than a bulk silicon device with the same thickness. 
On the other hand, it is known that the photocurrent density depends not only on the optical absorption but also on the surface recombination of excited charge carriers~\cite{Pankove000book}. 
Since an inverse woodpile crystal has a large surface area per unit cell compared to a thin homogeneous film, surface recombination requires further investigation 
to ascertain whether the enhanced absorption indeed leads to enhanced photovoltaic efficiy. 
Hence, a 3D inverse woodpile photonic crystal with a resonant cavity is an interesting candidate as absorbing medium to enhance the absorption of photons at multiple discrete frequencies in the visible range. 

\subsection{Fano resonances below the 3D band gap}
\label{subsect:FanoBelowBandGap}
\begin{figure}[tbp!]
\centering
\includegraphics[width=1\columnwidth]{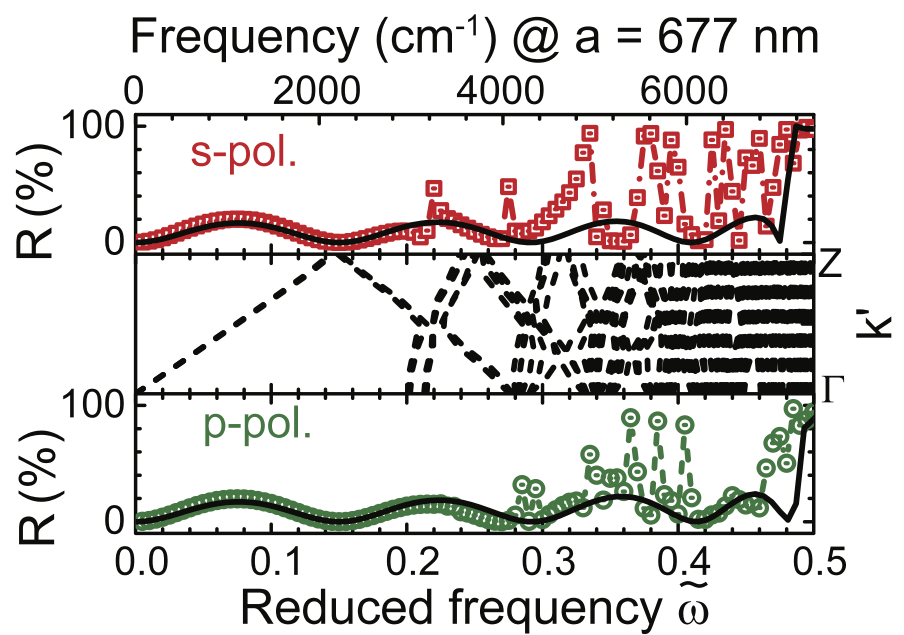}
\caption{Reflectivity below the 3D band gap for a 3D inverse woodpile photonic crystal with a point defect. 
The red squares in (i) and the green circles in (iii) are reflectivity spectra calculated at normal incidence for $s$ and $p$ polarizations, respectively. 
The corresponding band structure for the $\Gamma Z$ direction is shown in panel (ii). 
The wave vector is expressed as $k^{'} = (ka/2\pi)$. 
Black curves in (i) and (iii) indicate the reflectivity spectra for a perfect 3D inverse woodpile photonic crystal for $s$ and $p$ polarizations, respectively. 
Beyond $\tilde{\omega} = 0.21$, the reflectivity spectra for a crystal with a defect deviates increasingly from perfect crystal spectra.}
\label{fig:reflectivity_belowBandGap}
\end{figure}

The reflectivity spectra for a perfect 3D inverse woodpile photonic crystal reveal Fabry-P{\'e}rot fringes below the stop band as a result of interference between the front and back surfaces~\cite{Devashish2017PRB}. 
To investigate the effect of a point defect on these Fabry-P{\'e}rot fringes, we calculate the polarization-resolved ($s$ or $p$) reflectivity below $\tilde{\omega} = 0.50$ at normal incidence to the $3 \times 3 \times 3$ supercell of an inverse woodpile photonic crystal with and without a point defect, see Fig.~\ref{fig:reflectivity_belowBandGap}. 
Below $\tilde{\omega} = 0.21$ we observe in Fig.~\ref{fig:reflectivity_belowBandGap} that the spectra for a photonic crystal with two line defects matches very well with the one for a perfect photonic crystal. 
Beyond $\tilde{\omega} = 0.21$, however, several peaks appear. 
These peaks are narrow and sometimes have a reflectivity near $100\%$. 
The band structure in Fig.~\ref{fig:reflectivity_belowBandGap} (ii) reveal the characteristic band folding due to the supercell. 
The bands increase linearly up to $\tilde{\omega} = 0.12$ and then fold back up to $\tilde{\omega} = 0.21$. 
By comparing this band structure to the corresponding reflectivity spectra, we note that fringes for a photonic crystal with a defect match with the fringes for a perfect crystal only when the bands are in the linear dispersion regime $(\omega = c.k)$, that is, below $\tilde{\omega} = 0.21$. 
Therefore, these narrow peaks in the fringes correspond to frequency ranges of band folding. 

\begin{figure}[tbp!]
\centering
\includegraphics[width=1\columnwidth]{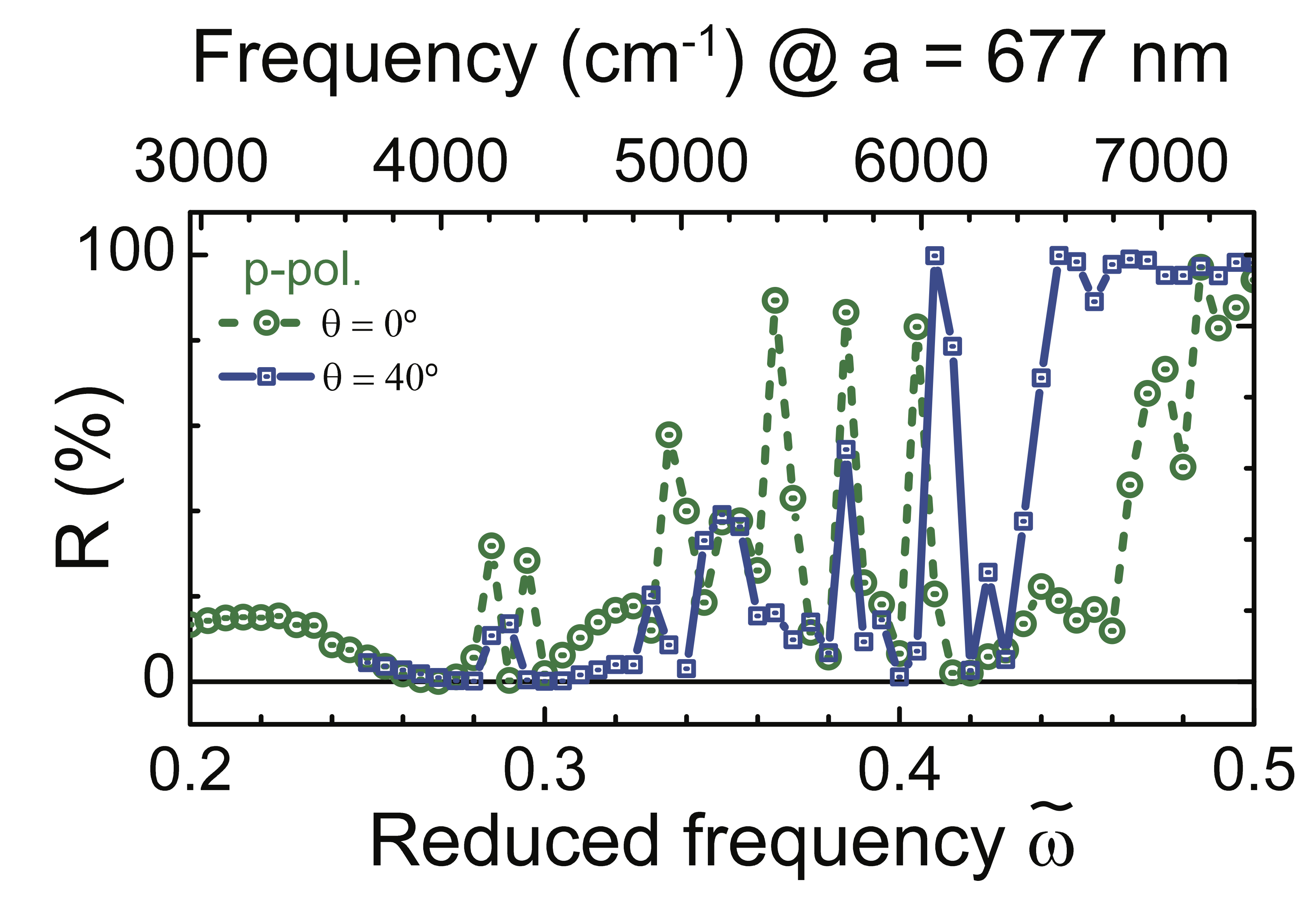}
\caption{Reflectivity below the 3D band gap for a 3D inverse woodpile photonic crystal with a point defect for $p$-polarized incident light. 
The green dashed and blue solid curves are the spectra for incident angles $\theta = 0\degree$ and $\theta = 40\degree$, respectively, off the normal in the $\Gamma Z$ direction.} 
\label{fig:reflectivity_belowBandGap_angle_P}
\end{figure}

To verify the confinement of light in real space at these peak frequencies, we investigate the angle-dependence of the peaks. 
Figure~\ref{fig:reflectivity_belowBandGap_angle_P} shows the reflectivity for a photonic crystal with a point defect between $\tilde{\omega} = 0.21$ and $\tilde{\omega} = 0.5$ at incident angles $0\degree$ and $40\degree$. 
We observe that there are peaks on top of the low-frequency Fabry-P{\'e}rot fringes for both incident angles. 
Most of these peaks vary with incident angle, unlike the cavity resonances in Section~\ref{subsect:WithinBandGap}, and thus these peaks are not cavity resonances.  

\begin{figure}[tbp!]
\centering
\includegraphics[width=1\columnwidth]{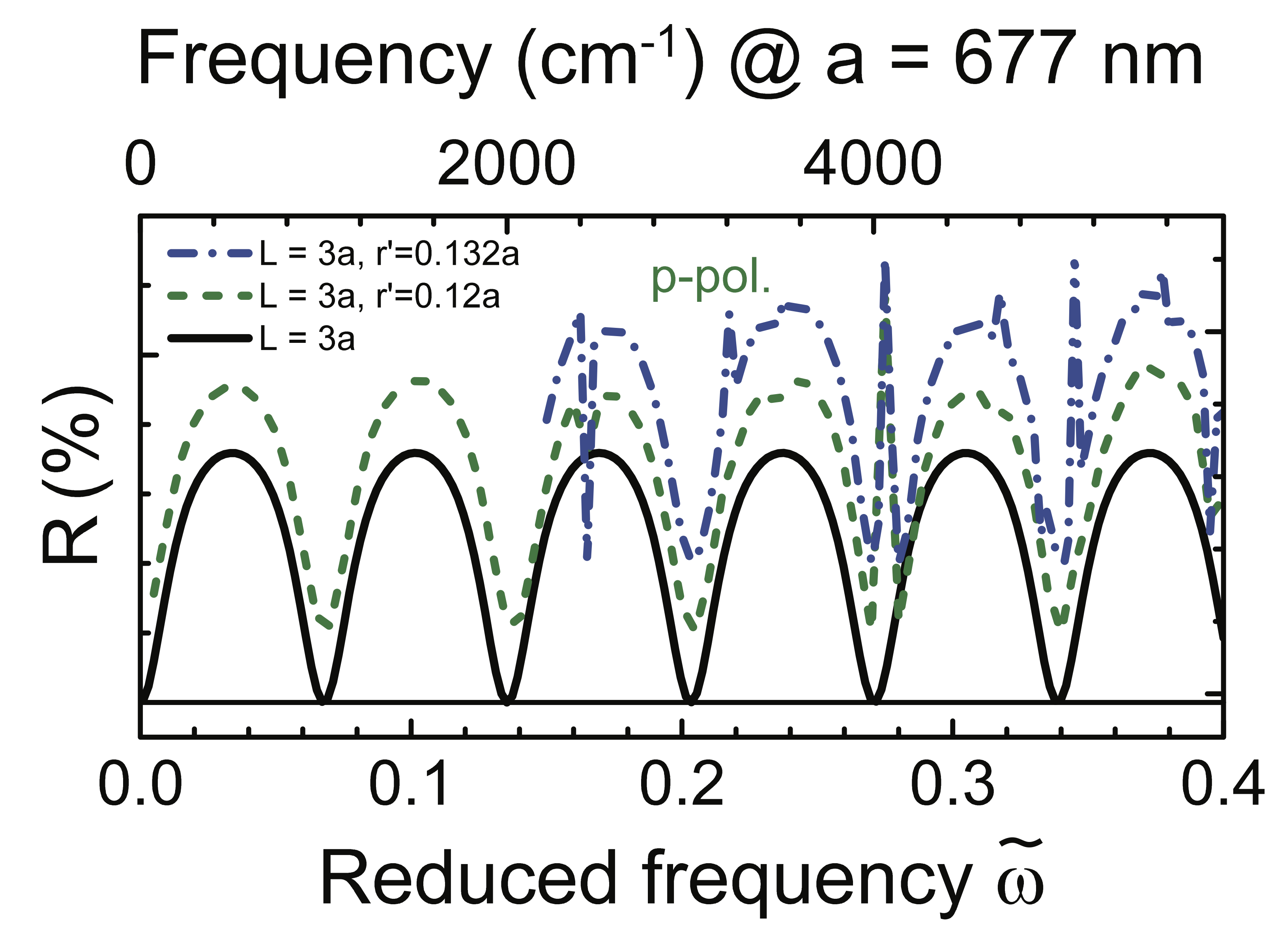}
\caption{Calculated reflectivity spectra for a thin dielectric film with thickness $L_{Z} = 3a$ for $p$ polarization. 
The film has dielectric permittivity $\epsilon = 12.1$. 
The black curve shows the reflectivity spectrum for a perfect thin film. 
Green dashed and blue dashed-dotted curves represent the reflectivity spectra for a thin homogeneous film with two orthogonal defect pores of radii $r'=0.12 a$ and $r'=0.132 a$, respectively. 
To compare spectra for different defect-pore radii, we use ordinate offsets of $20\%$ and $40\%$ in green dashed and blue dashed-dotted curves, respectively.} 
\label{fig:SiliconSlabwithDefect}
\end{figure}

To understand the origin of these peaks, we simplify the problem by studying a homogeneous thin silicon film ($L_{Z} = 3c = 900$ nm) with two orthogonal defect pores at the center (radii $r^{'} = 0.12a$), with the same computational cell as the inverse woodpile with the two defect pores. 
Figure~\ref{fig:SiliconSlabwithDefect} shows the reflectivity spectra for a thin film without and with two orthogonal defect pores. 
We observe Fabry-P{\'e}rot fringes below $\tilde{\omega} = 0.4$ that are due to the interference of the light with the two interfaces of the thin film. 
The reflectivity spectrum for a thin film with two orthogonal defect pores  reveals several peaks and troughs at discrete frequencies, \textit{e.g.}, a trough at $\tilde{\omega} = 0.18$ and a peak at $\tilde{\omega} = 0.27$. 
We note that a trough always occurs near the maxima and a peak always near the minima of a Fabry-P{\'e}rot fringe. 
Increasing radii of these defect pores changes amplitudes of existing peaks and troughs and new ones appear as well. 
We interpret this behavior as the electromagnetic interference of reflectivity from the continuum contribution of the film and from the discrete contribution of the defect pores. 
This interference leads to sharp asymmetric peaks, commonly referred as Fano resonances in solid-state and atomic physics~\cite{Fano1961PRB, Lukyanchuk2010NatMater, Fan2002PRB}. 

\begin{figure}[htbp!]
\centering
\includegraphics[width=1\columnwidth]{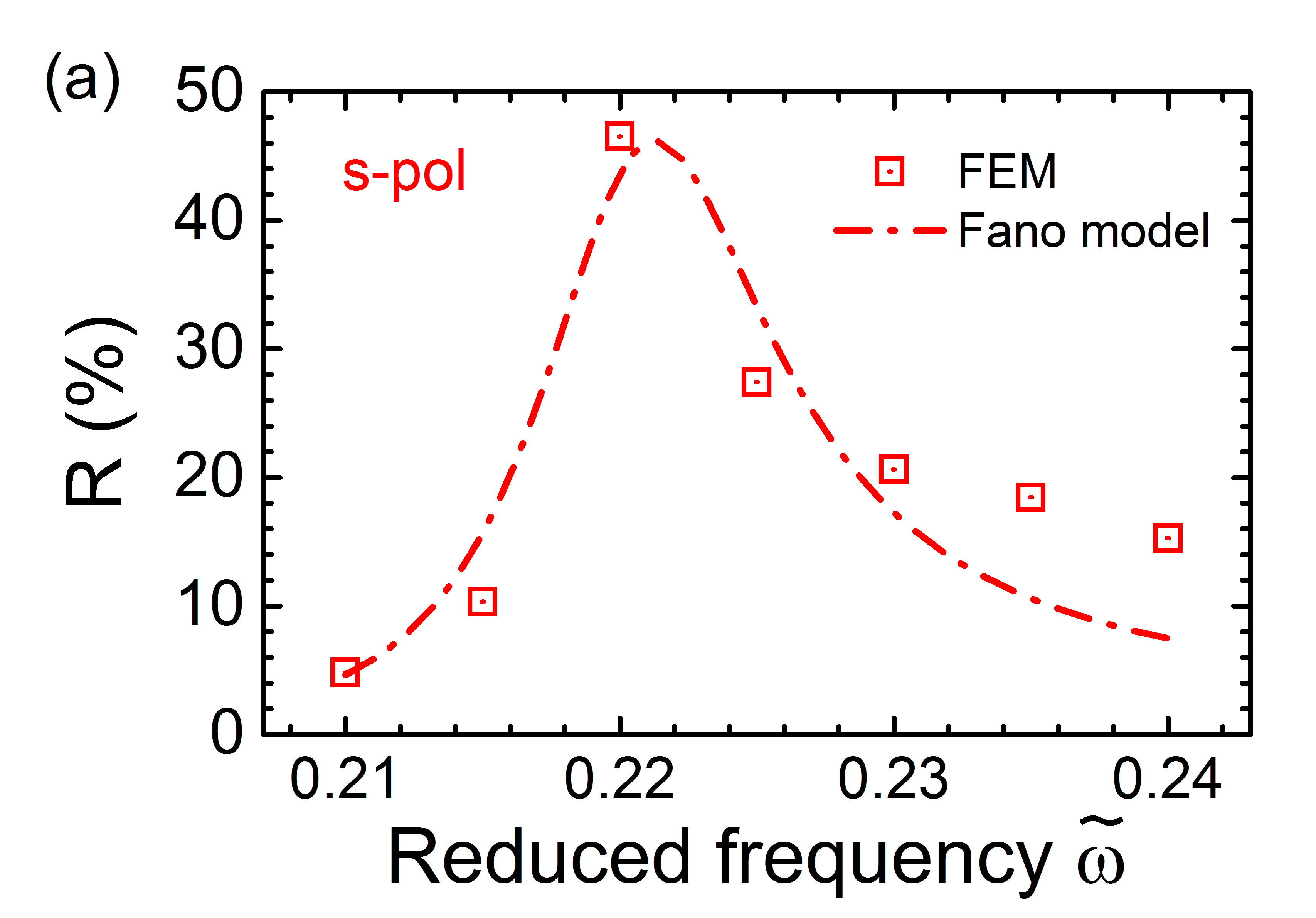}
\vspace{0.1cm}
\includegraphics[width=1\columnwidth]{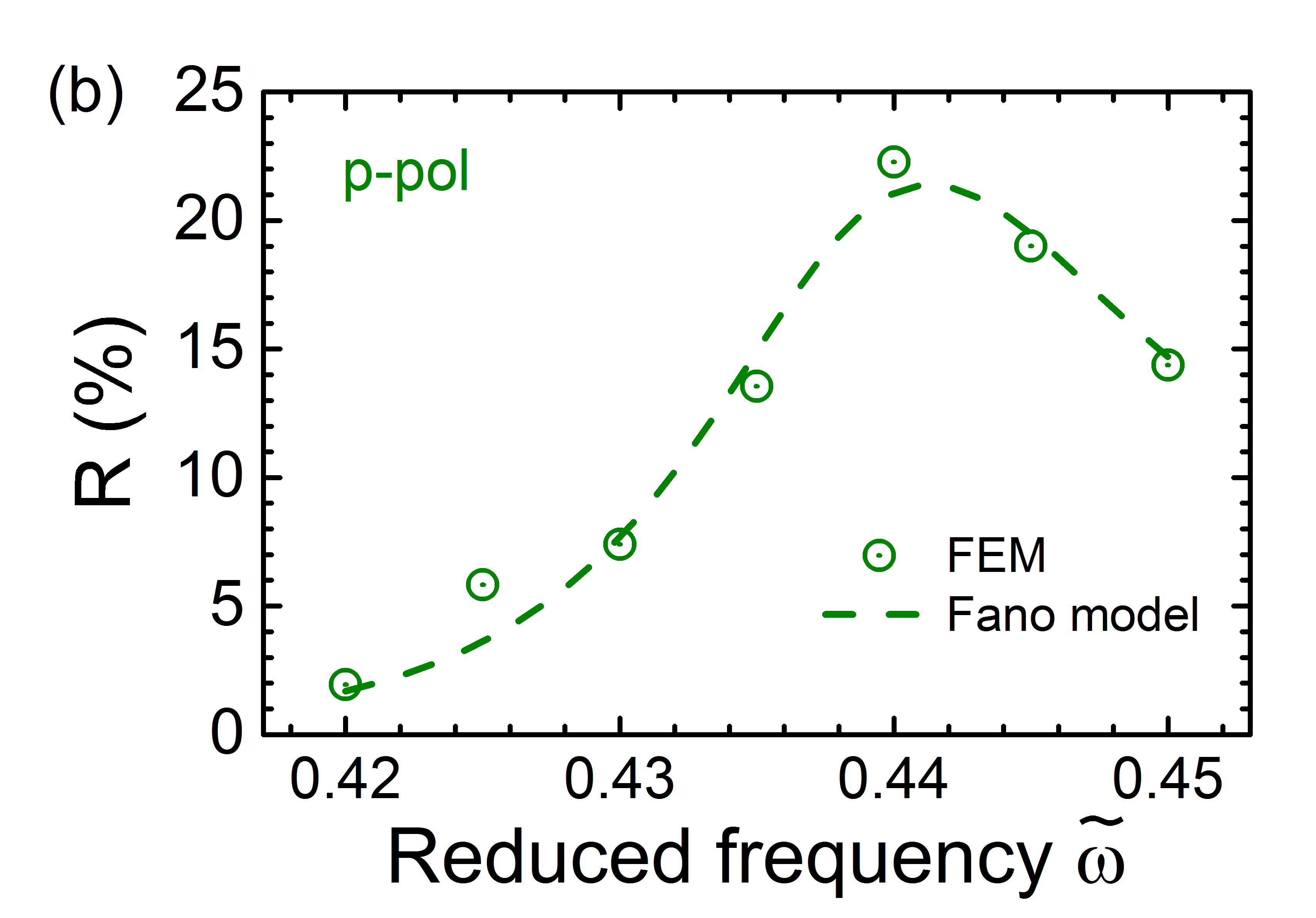}
\caption{Fano resonances below the 3D band gap for a 3D inverse woodpile photonic crystal with a cavity. 
The red squares in (a) and the green circles in (b) are zoomed-in reflectivity spectra calculated at normal incidence for $s$ and $p$ polarizations, respectively. 
The red dashed-dotted and green dashed curves denote the Fano resonance model~\cite{footnote:FanoFitting} fitted to the reflectivity peaks. 
}
\label{fig:Fano_belowBandGap}
\end{figure}

We now return to two orthogonal defect pores in an inverse woodpile. 
The incident plane wave is reflected from the photonic crystal for all frequencies and gets coupled to the resonant cavity only at discrete frequencies~\cite{Fan2002PRB,Vasco2013OptExpress}. 
Therefore, the electromagnetic interference between the continuum contribution of the light reflected by the photonic crystal and the discrete contribution of the cavity resonance gives rise to a Fano resonance~\cite{Fano1961PRB, Vasco2013OptExpress}, \textit{e.g.}, at $\tilde{\omega} = 0.22$ in the $s$-polarized spectrum in Fig.~\ref{fig:Fano_belowBandGap}(a). 
To confirm the reasoning above, we modeled the reflectivity data with the functional form of the Fano resonance ~\cite{footnote:FanoFitting, Fano1961PRB} in Fig.~\ref{fig:Fano_belowBandGap}(a,b).
When the continuum contribution and the discrete contribution to the interference are comparable, the Fano resonance has a characteristic sharp asymmetrical shape, \textit{e.g.}, at $\tilde{\omega} = 0.225$ for $s$ polarization in Fig.~\ref{fig:Fano_belowBandGap}(a) and at $\tilde{\omega} = 0.44$ for $p$ polarization in Fig.~\ref{fig:Fano_belowBandGap}(b).
Hence, these Fano resonances are angle-dependent and asymmetric in shape, unlike the cavity resonances that are angle-independent and have symmetric line shapes. 

\section{Discussion}
\subsection{Quadrupolar symmetry} 

Since Ref.~\cite{Woldering2014PRB} did not provide a symmetry-assignment of the cavity resonances, we propose an assignment here based on three main arguments. 
Firstly, the consistent observation of five resonances by three independent studies (Refs.~\cite{Woldering2014PRB,Hack2018} and the present work) strongly suggests that a cavity in an inverse woodpile photonic band gap crystal has eigenstates with quadrupole symmetry, since a quadrupole in electrodynamics has five components~\cite{Jackson1999Book}. 
Conversely, a cavity in an inverse woodpile crystal is analogous to d-orbitals in atomic solid-state physics, since d-orbitals have $5$ states ($d_{xy}$, $d_{xz}$, $d_{yz}$, $d_{x^2-y^2}$, $d_{z^2}$)~\cite{Ashcroft1976Book}. 

Secondly, the degeneracy of two bands at the $\Gamma$ high-symmetry point in Fig.~\ref{fig:ReflectivityAngle_S_P}(a,b) agrees with the occurrence of two degenerate bands out of five 3d-orbital bands of transition metal nickel at the conventional cubic $\Gamma$ points, see Fig.~10.6(a) in Ref.~\cite{Ashcroft1976Book}. 
Simultaneously, the degeneracy of the S1 and P1 bands both at $\Gamma$ and at about 2/3 along $\Gamma Z$ in Fig.~\ref{fig:ReflectivityAngle_S_P}(a,b) agrees with the occurrence of a degeneracy of two bands in the nickel case both at $\Gamma$ and at 0.9 of the $\Gamma K$ direction
(see Fig.~10.6(c) in Ref.~\cite{Ashcroft1976Book}), where the second point corresponds to 0.67 along $\Gamma Z$ in our tetragonal Brillouin zone~\cite{footnote:CubicLattice}.

Thirdly, Woldering \textit{et al.} presented (but did not interpret) in their Fig.~5 a cross-section of the energy density distributions in the $(x,z)$ plane of the crossing defect pores for a single resonance, namely  the 3rd one that corresponds to our P2 resonance. 
The cross-section reveals that $4$ sharp field maxima occur at $4$ sharp corners in the high-index dielectric material, with the same spatial distribution as the $4$ potential minima and maxima of an electric quadrupole~\cite{Jackson1999Book}. 
Since the field distribution pertains strictly to a \textit{single} resonance, one can immediately exclude the naive suggestion whether the $4$ field maxima pertain to a double dipole state. 
Taking these arguments together, we conclude that the resonances of the inverse woodpile cavity have quadrupolar symmetry and are the optical analogues of d-orbitals in solid-state physics. 

\subsection{Thickness-dependent quality factor}
In a finite crystal, it is common that perfect symmetries that pertain to infinitely extended crystals are disrupted~\cite{ChaikinBook}. 
Indeed, we have seen in the cross-correlation analysis above (Fig.~\ref{fig:EfieldCorrelation}) that for several finite-crystal resonances, a non-zero amount of other states has been admixed, on account of the observation of secondary maxima in the cross-correlation plots, \textit{e.g.}, the P1 resonance containing admixtures of the P2 and P4 resonances. 

\begin{figure}[tbp!]
\centering
\includegraphics[width=1.0\columnwidth]{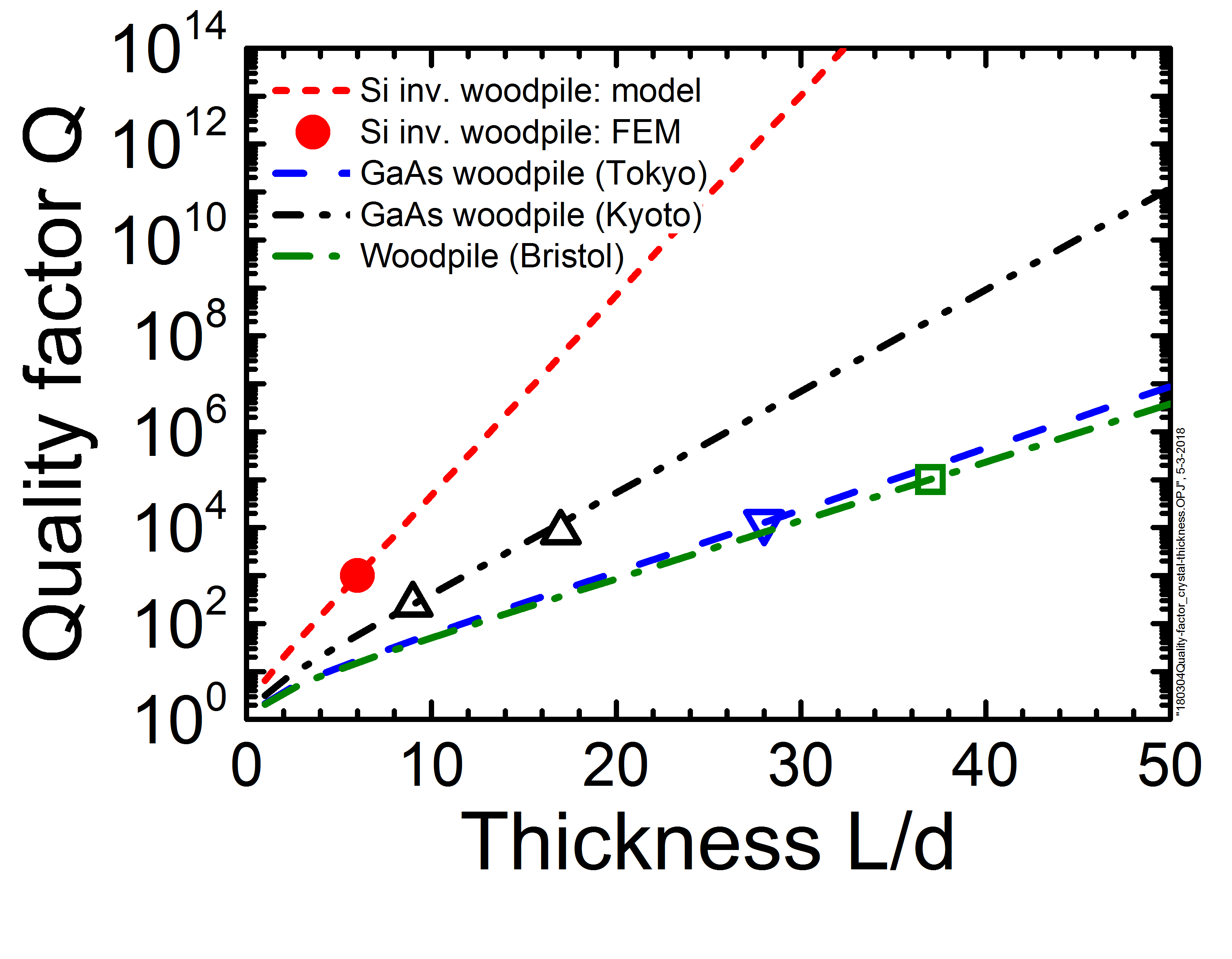}
\caption{Quality factor $Q$ of a photonic band gap cavity with finite support versus thickness of the embedding photonic band crystal. 
The thickness $L$ is reduced with the lattice parameter $d$. 
Red circle: our FEM result on Si inverse woodpile cavities; red dashed line: Eq.~(\ref{eq:QualityFactor_vs_Thickness}) with $S = 0.61$.
Black triangles: FDTD results on GaAs woodpile cavities~\cite{Ogawa2004Science}; black dashed-double-dotted line: Eq.~(\ref{eq:QualityFactor_vs_Thickness}) with $S = 0.31$.
Blue inverted triangle: experimental result on GaAs woodpile cavity~\cite{Tajiri2015APL}; blue dashed line: Eq.~(\ref{eq:QualityFactor_vs_Thickness}) with $S = 0.19$.
Green square: FDTD result on woodpile structure~\cite{Taverne2015JOSAB}; green dashed-dotted line: Eq.~(\ref{eq:QualityFactor_vs_Thickness}) with $S = 0.18$.} 
\label{fig:Q_vs_thickness}
\end{figure}

Let us put the quality factors above in context with other 3D photonic band cavities~\cite{Ogawa2004Science,Tajiri2015APL,Taverne2015JOSAB}.
Since all results to date pertain to photonic band gap crystals with different thicknesses, different high-index backbones, and different crystal structures, we derive an analytic model. 
Therefore, we invoke a simplified one-dimensional (1D) model for a photonic band gap cavity, that corresponds to a microcavity consisting of two Bragg stacks surrounding a central defect layer (\textit{cf.} Ref.~\cite{Gerard2003TAP}. 
For such a cavity, we derive (see Appendix~\ref{sect:Q_versus_thickness}) that the quality factor $Q$ increases exponentially with the thickness $L$ of the cavity structure 
\begin{align}
\textrm{Q} = \pi.\exp \left(\pi \frac{L}{2d}\frac{\Delta \tilde{\omega}_{s}}{\tilde{\omega}_{s}}\right),
\label{eq:QualityFactor_vs_Thickness}
\end{align}
with $d$ the lattice spacing of the photonic crystal, and $\Delta \tilde{\omega}_{s}$ the band width and $\tilde{\omega}_{s}$ the central frequency of the stop band. 
In Fig.~\ref{fig:Q_vs_thickness}, we plot the available data for all photonic band gap cavities. 
We model each data set with Eq.~(\ref{eq:QualityFactor_vs_Thickness}) where the only adjustable parameter is the ratio $S = \Delta \tilde{\omega}_{s}/\tilde{\omega}_{s}$ that is taken to be equal to the photonic interaction strength $S$ of a photonic crystal~\cite{Vos2015Book}, and that is inversely proportional to the Bragg length that gauges the typical length scale for interference in a photonic stop gap. 

Figure~\ref{fig:Q_vs_thickness} shows that the computed quality factor $Q = 10^3$ of the best confined resonances (P1, P2, P3) of the Si inverse woodpile structure occurs for a thickness of $L/d = 6$ lattice planes~\cite{footnote:LatticePlane}. 
If we model this data point with Eq.~(\ref{eq:QualityFactor_vs_Thickness}), we find $S = 0.61$, which is considerably larger than the relative width of the dominant $\Gamma-X$ or $\Gamma-Z$ stop bands ($\Delta \tilde{\omega}_{s}/\tilde{\omega}_{s} = 0.2$) and than the width of the $\Gamma-Y$ stop band ($\Delta \tilde{\omega}_{s}/\tilde{\omega}_{s} = 0.3$). 
We surmise that the effective photonic strength is rather large, since the inverse woodpile structure has a topology and connectivity that are very amenable to photonic gap formation, see Ref.~\cite{Economou1993PRB}. 
Ogawa \textit{et al.} report quality factors computed by FDTD simulations for cavities in GaAs woodpile crystals with two different thicknesses ($L/d = 9$ and $17$)~\cite{Ogawa2004Science}. 
These results are described by Eq.~(\ref{eq:QualityFactor_vs_Thickness}) with $S = 0.31$, which matches well with the stop band widths in band structures. 
Tajiri \textit{et al.} report an experimentally measured $Q = 12800$ for a GaAs woodpile crystal with a thickness $L/d = 9$. 
This results is described by Eq.~(\ref{eq:QualityFactor_vs_Thickness}) with $S = 0.19$, in reasonable agreement with the stop band widths in the band structures. 
Since in general computed quality factors are greater than measured ones (see, e.g., Ref~\cite{Ogawa2004Science}), the photonic strength derived from computed results on the perfect structure is expected to also be much greater. 
Taverne \textit{et al.} report the quality factor in a woodpile structure from FDTD simulations, and obtain $Q = 10^5$ for $L/d = 37$ layers~\cite{Taverne2015JOSAB}. 
Interpreting this result with Eq.~(\ref{eq:QualityFactor_vs_Thickness}) yields $S = 0.18$, in agreement with the stop band widths in the band structures.

\section{Conclusion}
We have numerically studied the reflectivity and the absorption of a resonant cavity in a three-dimensional photonic crystal with finite support. 
We employed the finite element method to study crystals with the cubic diamond-like inverse woodpile structure with a high-index backbone having a dielectric function similar to silicon. 
The point defect functioning as a cavity is formed in the proximal region of two orthogonal pores with a radius that differs from all others in the bulk of the crystal. 
By comparing defect bands in the band structure for an infinite crystal with resonances in the reflectivity spectra for a finite crystal, we identify cavity resonances and their field patterns. 
Out of five observed cavity resonances, one is $s$-polarized and four are $p$-polarized. 
These cavity resonances are angle-independent, indicating a strong confinement of light in the crystal slab. 
The P1, P2, and P4 resonances reveal normal behavior with single cross-correlation peaks (between field distributions) and single reflectivity resonances. 
The P3 and S1 resonances in finite crystals reveal an intriguing splitting into 2 sub-resonances. 
We find large energy enhancement at cavity resonances, \textit{i.e.}, up to $\eta_{E} = 2400$ times the incident energy and up to $\eta_{E} = 1200$ times the energy at a lower frequency. 
Our results indicate that 3D photonic band gap crystals with resonant cavities are potential candidates for the absorbing medium of a solar cell in order to enhance the photovoltaic efficiency and reduce the weight of the absorbing component by nearly $80\%$. 
Fano resonances are observed below the band gap due to the electromagnetic interference between the discrete contribution of the fundamental cavity mode and the continuum contribution of the light scattered by the photonic crystal. 
Our study indicates that the five eigenstates of our 3D photonic band gap cavity have quadrupolar symmetry, in analogy to d-like orbitals in solid-state physics. 
We conclude that inverse woodpile cavities have intriguing potential to applications in optical sensing and photovoltaics. 


\section{Acknowledgments} 
\label{sect:Acknowledgements} 
It is a pleasure to thank Bart van Tiggelen (CNRS, Grenoble), Jean-Michel G{\'e}rard (CEA, Grenoble), Sjoerd Hack, Ad Lagendijk, Allard Mosk (Utrecht), Pepijn Pinkse, Ravitej Uppu, and Satoshi Iwamoto, Takeyoshi Tajiri (Tokyo) and Shun Takahashi (Kyoto) for stimulating discussions, and Manashee Adhikary for help. 
This research is supported by the Shell-NWO/FOM programme ``Computational Sciences for Energy Research" (CSER), by the FOM/NWO programme ``Stirring of light!," the STW/NWO-Perspectief program ``Free-form scattering optics", the  ``Descartes-Huygens" prize of the French Academy of Sciences, and the MESA$^{+}$ Institute for Nanotechnology section Applied Nanophotonics (ANP). 


\appendix

\section{Comparison of the numerical solvers}
\label{sect:PhotonicBandstructure}

\begin{figure}[bp!]
\centering
\includegraphics[width=1\columnwidth]{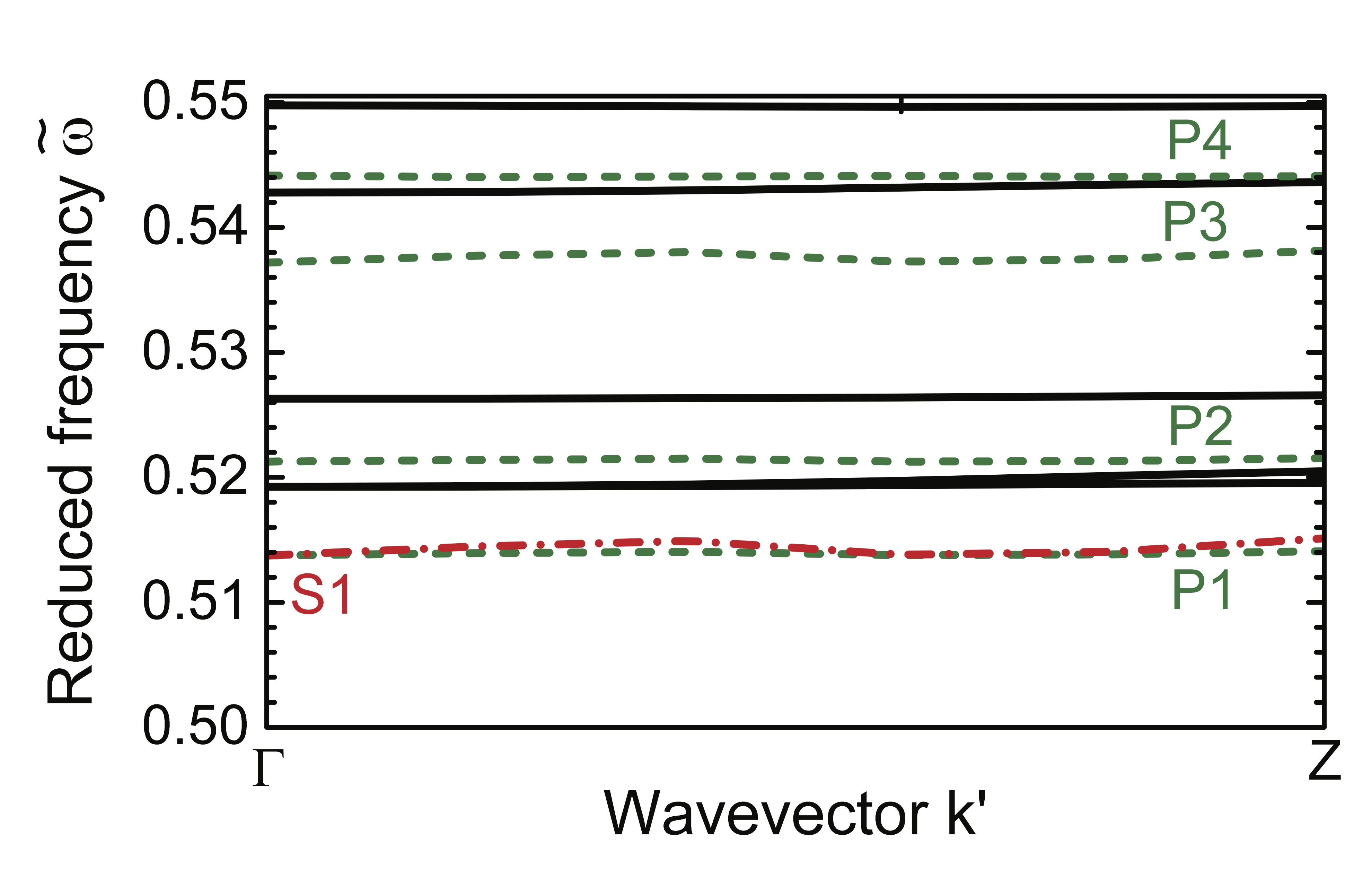}
\caption{Band structures in the $\Gamma Z$ direction calculated with the finite-element method (FEM) solver (red dashed-dotted line and green dashed lines) and with the MPB plane-wave expansion (PWE) method (black lines). 
Due to higher spatial resolution, the bands calculated by FEM are shifted to lower frequencies compared to those calculated by MPB. 
S1 is the $s$-polarized cavity resonance, 
and P1, P2, P3, and P4 are the four $p$-polarized cavity resonances in the photonic band gap.}
\label{fig:bandstructure_pwe_fem}
\end{figure}

To compute the photonic band structure for the supercell of a 3D inverse woodpile photonic crystal with a point defect, Ref.~\cite{Woldering2014PRB} employs a plane-wave expansion (PWE) method eigenvalue solver, whereas we employ the COMSOL finite-element method (FEM)~\cite{COMSOLMultiphysics} eigenvalue solver~\cite{Johnson2001OptExpress}. 
We compute the polarization-resolved (both $s$ and $p$) band structures. 
Figure~\ref{fig:bandstructure_pwe_fem} shows the bands between $\tilde{\omega} = 0.51$ and $\tilde{\omega} = 0.56$ obtained using both the FEM solver and the PWE method.  
We observe that there are five isolated and nearly dispersionless bands obtained using both methods. 
Out of these five bands, the FEM solver gives one $s$-polarized band, namely S1, and four $p$-polarized bands, namely P1, P2, P3, and P4. 
We note that all bands obtained using the FEM occur at lower frequencies compared to the corresponding bands obtained using the PWE method. 

\begin{figure}[tbp!]
\centering
\includegraphics[width=1\columnwidth]{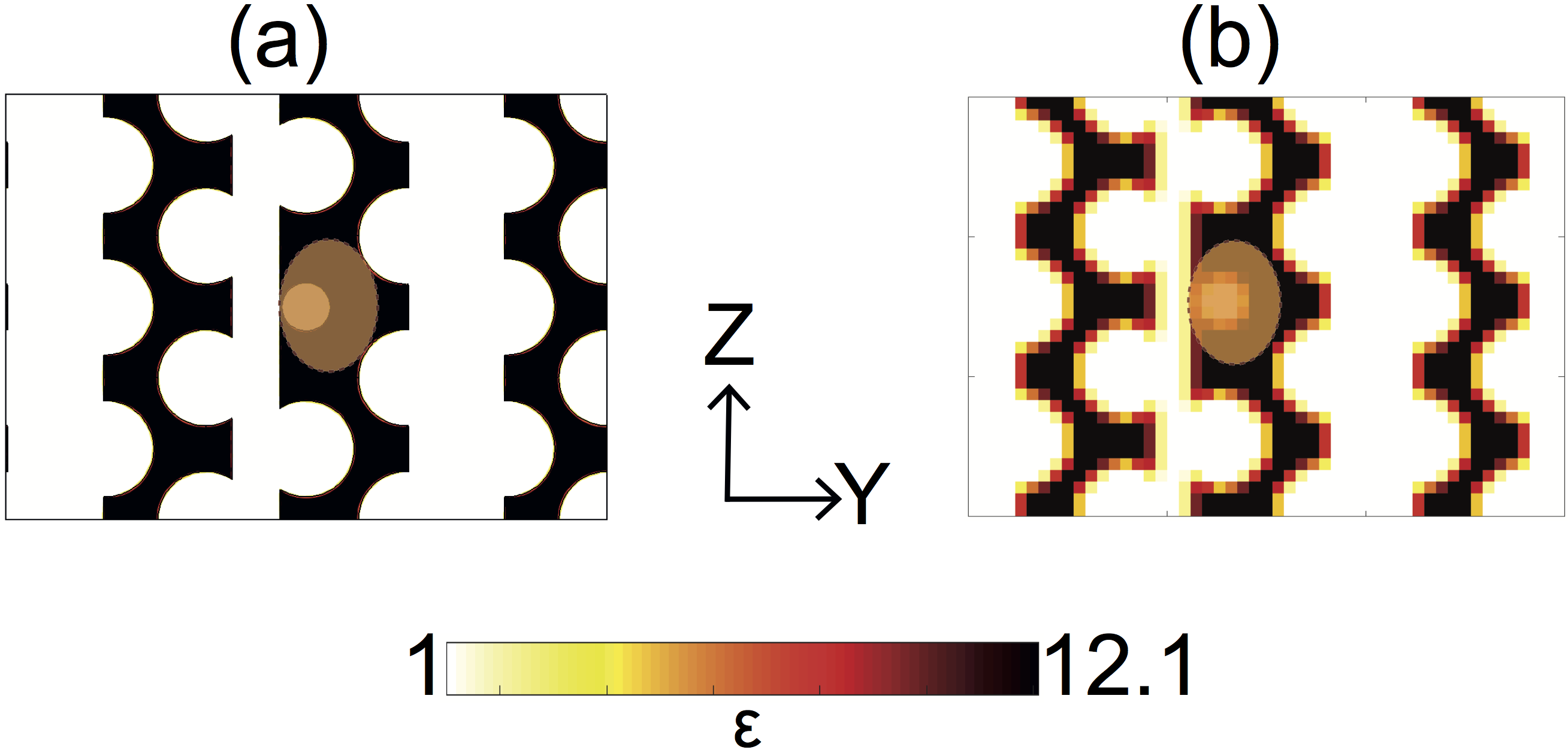}
\caption{Cross sections of the dielectric-permittivity distribution ($\epsilon$) along the $YZ$ plane of the $3 \times 3 \times 3$ super cell of a 3D inverse woodpile photonic crystal with a point defect, using (a) COMSOL finite-element method and (b) MPB plane-wave expansion method with a grid resolution of $24 \times 34 \times 24$.
The black color represents the high-index backbone of the crystal and the white color represents air. 
Two orthogonal defect pores result in a region with an excess of high-index backbone, as highlighted by orange ellipses.
} 
\label{fig:epsilon_yz}
\end{figure}

In order to characterize this frequency shift between the FEM and the PWE results, we compare cross sections of the dielectric-permittivity distribution of the supercell structure obtained from both methods in Fig.~\ref{fig:epsilon_yz}. 
We take the cross section through the center of the cavity and parallel to the $YZ$ plane. 
We observe that the curved boundaries between the air and the high-index backbone material are smoother in Fig.~\ref{fig:epsilon_yz} (a) compared to Fig.~\ref{fig:epsilon_yz} (b). 
This difference is more pronounced for the sharp interface surrounding the point defect. 
Compared to the PWE solver, we use a smaller element size in the FEM solver to subdivide the computational domain and hence the sharp interfaces and the curved boundaries are better approximated. 
Ref.~\cite{Woldering2014PRB} also reports in Appendix A that the resonance bands shift to lower frequencies with a higher spatial resolution. 
Therefore, we conclude that the frequency shift between the bands obtained using the two numerical methods is due to the differences in the spatial resolution, resulting into different dielectric-permittivity distributions. 

The computational time using the FEM solver is $2 \times$ longer than the PWE method. 
In order to minimize the computational time, we subdivide the frequency regime into two ranges: below the 3D band gap and the 3D band gap. 
Since there are no isolated resonance bands below the 3D band gap, we do not need the calculation to have the FEM spatial resolution. 
Hence, we employ the faster option of the PWE method to calculate the photonic band structure below the band gap. 
However, we employ a spatial resolution of $24 \times 34 \times 24$, which is a $2 \times 2 \times 2$ times greater 3D spatial resolution than in Ref.~\cite{Woldering2014PRB}.
Since we explicitly aim at identifying isolated cavity resonances, we compute the photonic band structure inside the 3D band gap using the FEM solver, which is the same numerical method used for the reflectivity calculations.   

\section{Mesh convergence}
\label{sect:MeshConvergence}

\begin{figure}[tbp!]
\centering
\includegraphics[width=1\columnwidth]{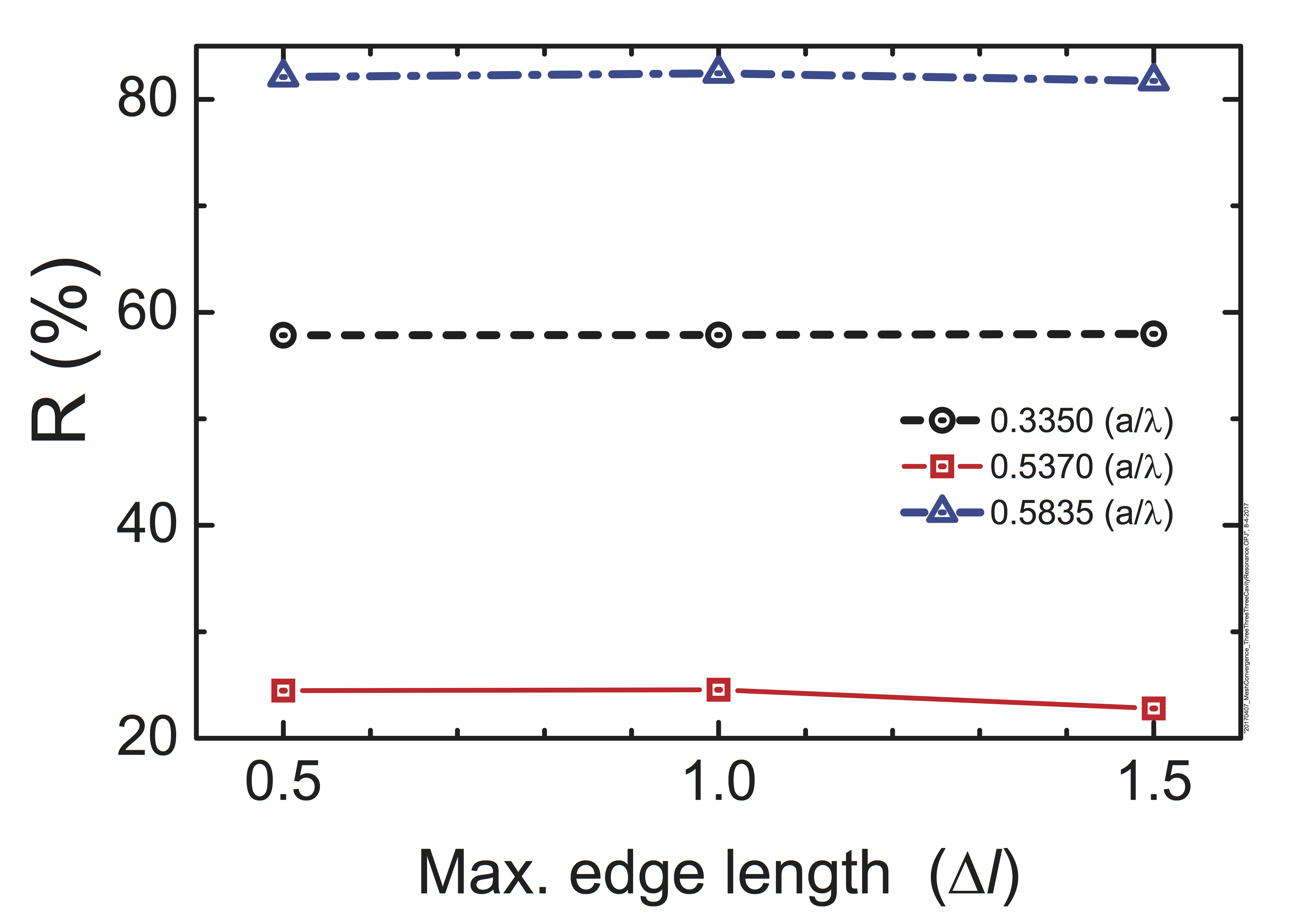}
\caption{Mesh convergence analysis of the finite-element method for the calculation of reflectivity spectra for a 3D inverse woodpile photonic crystal with a point defect. 
Black circles and blue triangles pertain to frequencies below and near the upper edge of the 3D band gap, respectively. 
Red squares denote the frequency at the P3 cavity resonance. 
Black dashed, blue dashed-dotted, and red solid curves are guides to the eye showing modulations in the reflectivity with the varying edge length of a tetrahedron.} 
\label{fig:MeshConvergence_P}
\end{figure}
In this study, we use tetrahedra to subdivide the 3D computational domain in the finite element method. 
To determine the edge length of the tetrahedra to completely mesh the complex geometry, we investigate the mesh convergence of the reflectivity results. 
We perform reflectivity calculations using upper limits of $\triangle l \leqslant \frac{\lambda_{0}}{4\sqrt{\epsilon}}$, $\triangle l \leqslant \frac{\lambda_{0}}{8\sqrt{\epsilon}}$, and $\triangle l \leqslant \frac{\lambda_{0}}{12\sqrt{\epsilon}}$ to the edge length $\triangle l$ on any tetrahedra, with $\lambda_{0}$ the shortest wavelength of the incident plane waves in vacuum. 
Figure~\ref{fig:MeshConvergence_P} shows the reflectivity at frequencies below, inside, and near the upper edge of the 3D band gap. 
These reflectivity values change less than $\sim 0.1\%$ with the maximum edge length. 
From the nearly constant results of these three mesh resolutions, we conclude the quantitative convergence of our calculated reflectivity spectra.  
These three mesh resolutions take 4300 s, 5330 s, and 11020 s computation time on the Serendipity cluster~\cite{Serendipity}. 
Therefore, to keep the computational time for many frequencies tractable while maintaining the quantitative convergence of the results, we set an upper limit of $\triangle l \leqslant \frac{\lambda_{0}}{8\sqrt{\epsilon}}$ to the edge length of any tetrahedra used in the finite-element mesh.

\section{Frequency convergence}
\label{sect:FrequencyConvergence}

\begin{figure}[!htbp]
\centering
\includegraphics[width=0.85\columnwidth]{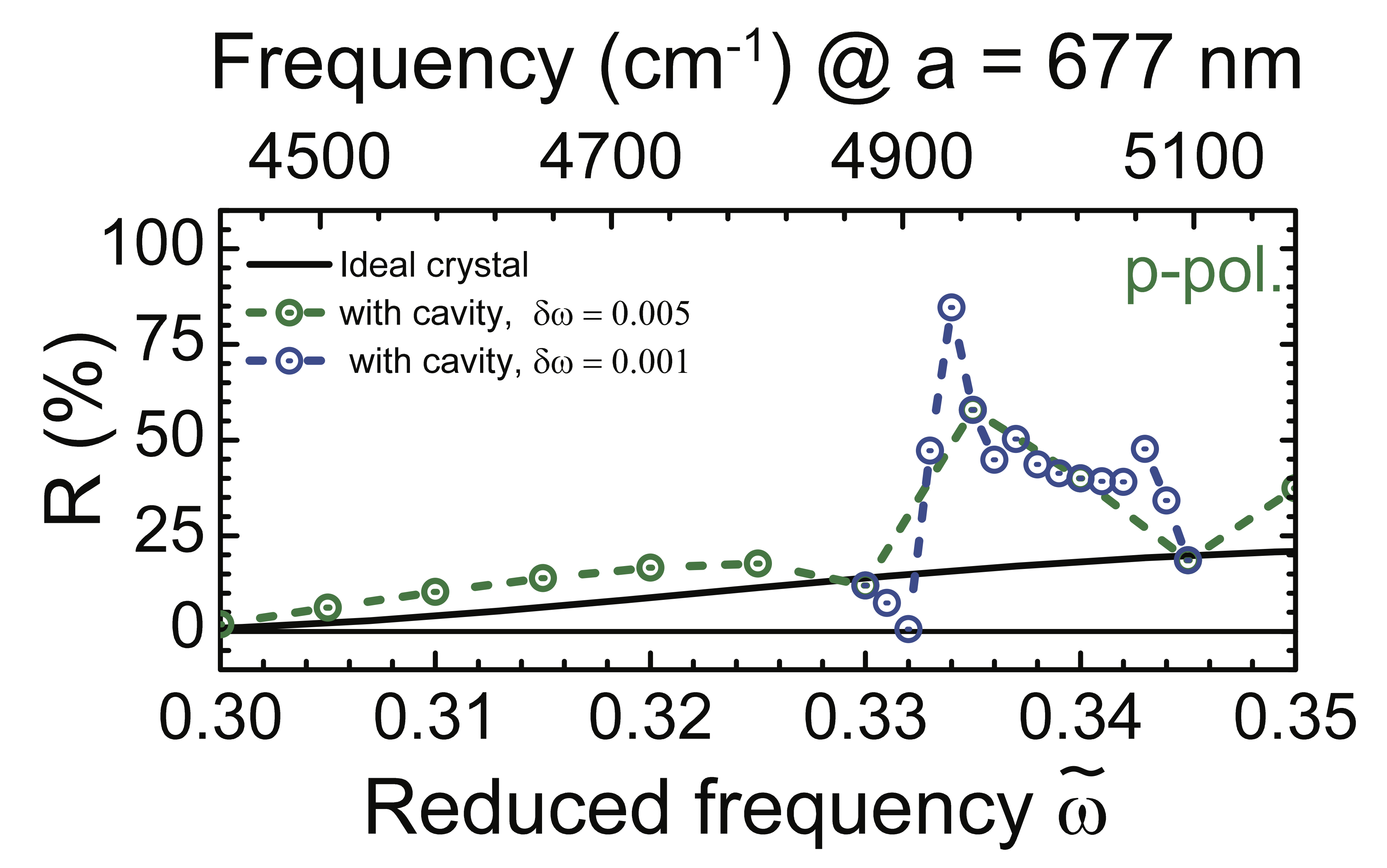}\\
\vspace{0.3cm}
\includegraphics[width=0.85\columnwidth]{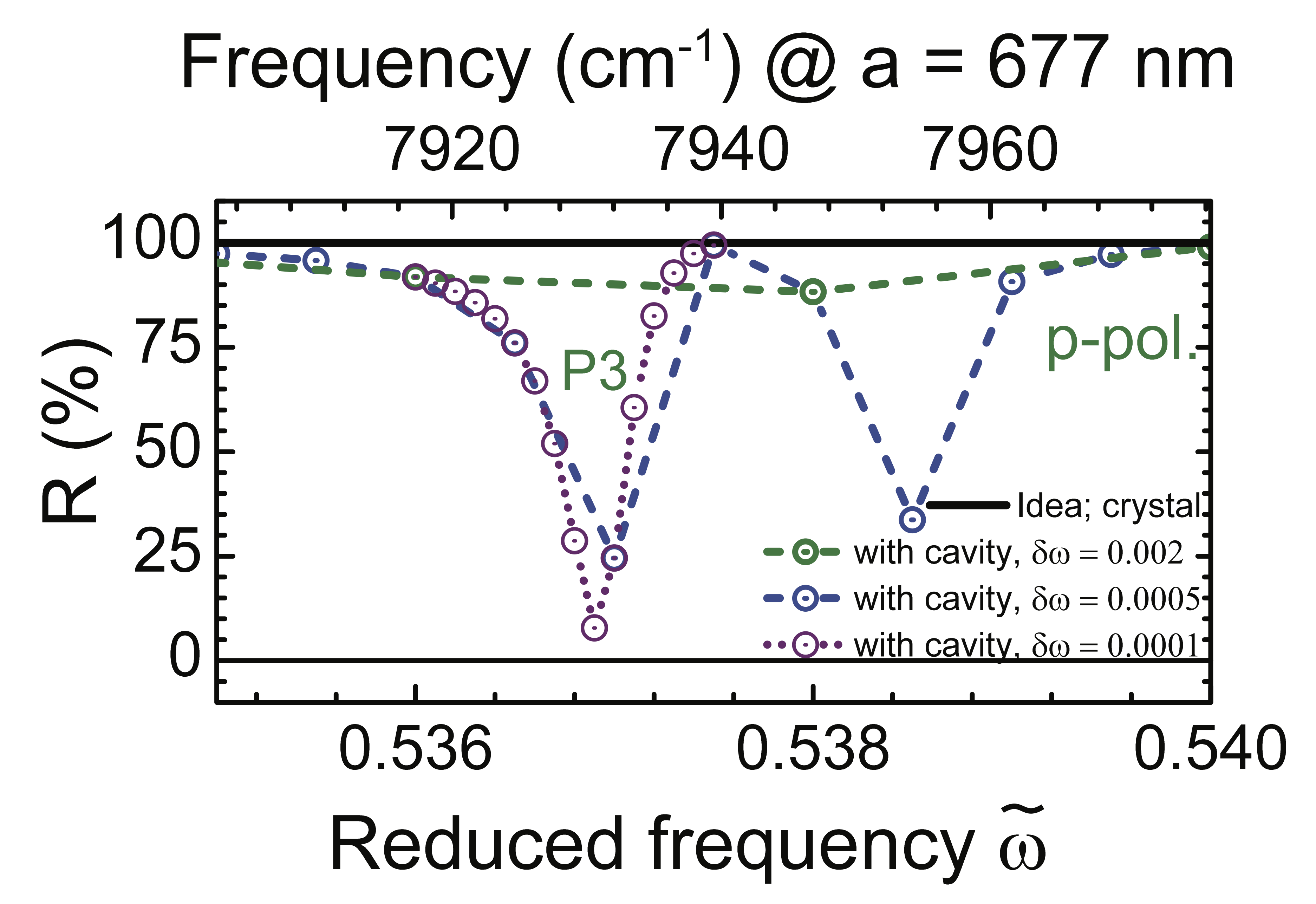}\\
\vspace{0.3cm}
\includegraphics[width=0.85\columnwidth]{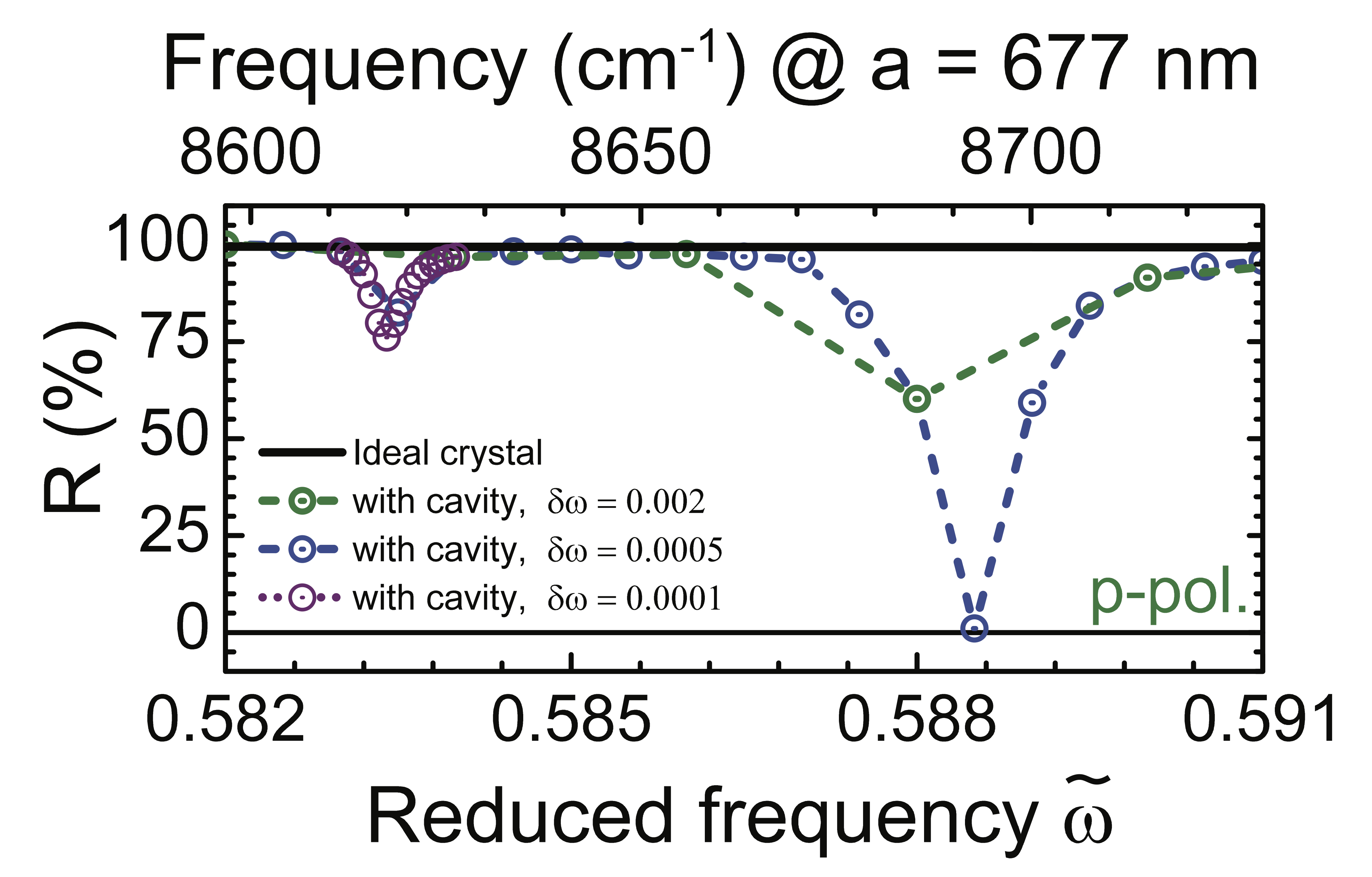}
\caption{Reflectivity spectra for a 3D inverse woodpile photonic crystal with a point defect for $p$-polarized light at normal incidence in the $\Gamma Z$ direction. 
(a) Fano resonance below the band gap, (b) P3 cavity resonance inside the band gap, and (c) resonance near the upper edge of the band gap. 
Green dashed, blue dashed, and purple dashed curves are results calculated with frequency resolutions $\delta\tilde{\omega} = 0.002$, $0.0005$, and $0.0001$, respectively.
The black solid line indicates the reflectivity for a perfect crystal without defect.} 
\label{fig:Frequency_Convergence}
\end{figure}

The reflectivity troughs corresponding to the cavity resonances have bandwidths as narrow as $\triangle \tilde{\omega} = 0.0005$. 
Thus, a calculation performed for an insufficient number of discrete frequencies will not detect these reflectivity resonances. 
Moreover, the calculation may not show the actual minima of a given trough due to saturation. 
Therefore, we perform the frequency convergence analysis to determine the appropriate frequency resolution to detect these resonance troughs.  
We define the frequency resolution as the spacing between two adjacent frequencies.
We choose three frequency regimes: below the band gap between $\tilde{\omega} = 0.30$ and $\tilde{\omega} = 0.35$, inside the band gap between $\tilde{\omega} = 0.53$ and $\tilde{\omega} = 0.54$, and near the upper edge of the band gap between $\tilde{\omega} = 0.58$ and $\tilde{\omega} = 0.59$. 

Figure~\ref{fig:Frequency_Convergence}(a) shows the reflectivity spectra between $\tilde{\omega} = 0.30$ and $\tilde{\omega} = 0.35$ below the band gap. 
A comparison between the spectra for a perfect inverse woodpile and an inverse woodpile with a point defect reveals a Fano resonance at $\tilde{\omega} = 0.335$, as previously shown in Section~\ref{subsect:FanoBelowBandGap}. 
We observe that the maximum of this Fano resonance increass with frequency resolution. 
Also, a new Fano resonance appears at $\tilde{\omega} = 0.335$ at higher frequency resolution. 
Since we performed calculations using $\delta\tilde{\omega} = 0.005$ for the frequency range below the 3D band gap, there could be more Fano resonances than the ones shown in Fig.~\ref{fig:reflectivity_belowBandGap}.

Figure~\ref{fig:Frequency_Convergence}(b) shows the reflectivity spectra inside the band gap between $\tilde{\omega} = 0.53$ and $\tilde{\omega} = 0.54$. 
We observe that P3 cavity resonance troughs at $\tilde{\omega} = 0.536$ and $\tilde{\omega} = 0.538$ are detected only at frequency resolutions $\delta\tilde{\omega} = 0.0005$ and $0.0001$. 
Since we employed a frequency resolution $\delta\tilde{\omega} = 0.0005$ for all calculations inside the 3D band gap, we have successfully detected all possible troughs.
However, we note that the minima of the trough with $\tilde{\omega} = 0.536$ changes around $25\%$ after increasing the frequency resolution by 5 times. 
Therefore, the observed cavity resonances may show even lower minimum reflectivity at a higher frequency resolution than our present calculations.

Figure~\ref{fig:Frequency_Convergence}(c) shows the reflectivity spectra near the upper edge of the band gap between $\tilde{\omega} = 0.582$ and $\tilde{\omega} = 0.591$. 
We see two troughs at $\tilde{\omega} = 0.583$ and $\tilde{\omega} = 0.588$. 
We observe that the trough at $\tilde{\omega} = 0.583$ is invariant with frequency resolution, whereas the minimum value for the trough at $\tilde{\omega} = 0.588$ changes. 
Therefore, we surmise that the trough at $\tilde{\omega} = 0.583$ is a numerical speckle due to the finite sized calculations whereas the trough at $\tilde{\omega} = 0.588$ corresponds to one of the resonance bands near the upper edge of the band gap. 

\section{Quality factor versus crystal thickness}
\label{sect:Q_versus_thickness}
We derive an analytic expression for the quality factor $Q$ as a function of the thickness $L$ of the photonic crystal slab. 
We invoke a simplified one-dimensional (1D) model for a photonic band gap cavity, that corresponds to a Fabry-Perot microcavity consisting of two Bragg mirrors surrounding a central defect layer (\textit{cf.} Ref.~\cite{Gerard2003TAP}. 
For a 1D planar Fabry-Perot microcavity, the finesse $F$ is expressed in terms of the transmission $T$ of the mirrors as~\cite{Demtroder000book}
\begin{align}
F = m.Q = \pi \frac{\sqrt{(1-T)}}{T},
\label{eq:Finesse_vs_Transmission}
\end{align}
where the finesse is equal to the $m$th order resonance times the quality factor $Q$. 
The Bragg mirrors have a photonic stop gap, where the transmission decreases exponentially with thickness $L$ 
\begin{align}
T = \exp(-L/\ell_B). 
\label{eq:Transmission_vs_Thickness}
\end{align}
Here, $\ell_B$ is the characteristic Bragg length that can be expressed in terms of the photonic strength (relative stop gap width ($\Delta \omega_{s} /\omega_{s}$), with $\Delta \tilde{\omega}_{s}$ the band width and $\tilde{\omega}_{s}$ the central frequency of the stop band) as 
\begin{align}
\ell_B = \frac{2d}{\pi} \frac{\omega_{s}}{\Delta \omega_{s}}, 
\label{eq:Bragg-length_vs_Phot-strength}
\end{align}
with $d$ the lattice spacing of the Bragg mirrors. 
Taking Eqs.~\ref{eq:Finesse_vs_Transmission},~\ref{eq:Transmission_vs_Thickness},~\ref{eq:Bragg-length_vs_Phot-strength} together, considering that a typical microcavity resonance is of a low order ($m \simeq 1$), and considering that in the stop gap the transmission very small ($T << 1$), we arrive at the expression for the quality factor $Q$ increasing exponentially with the thickness $L$ of the cavity structure 
\begin{align}
\textrm{Q} = \pi.\exp \left(\pi \frac{L}{2d}\frac{\Delta \tilde{\omega}_{s}}{\tilde{\omega}_{s}}\right). 
\label{eq:QualityFactor_vs_Thickness_Appendix}
\end{align}



\begin{thebibliography}{}
\bibitem{footnote:ReducedFrequency}
In this paper, we express frequency as a reduced frequency $\tilde{\omega} = \omega a / (2 \pi c'))$, with $\omega$ the frequency, $a$ the lattice parameter, $c'$ the speed of light (not to be confused with the lattice parameter $c$). 
In this definition, the reduced frequency $\tilde{\omega}$ equals $(a/\lambda)$.
\bibitem{Serendipity}
"Serendipity" is a high performance computing cluster whose main features are 16 Dell Power edge R430 servers with each 2x Intel Xeon E5-2698 processors at 2.2GHz, with in total 640 cores and about 3.6 terabyte memory, connected through Infiniband. 
The head node is a Dell Power Edge R730.
\bibitem{footnote:NormEField}
Since the spatial distribution of the electric field is computational memory intensive, the norm of electric field is employed for the correlation calculation.
\bibitem{footnote:P3Bandwidth}
Since the double reflectivity resonance may be the result of a (currently) unknown coupling, it is uncertain whether the widths of either resonance can be attributed to a true resonance width and thus a quality factor. 
In absence of further information, we derive the bandwidths of P3$^{-}$ and P$^{+}$ reflectivity resonances to be $\triangle \tilde{\omega}_{-}  = 0.00053$ and $\triangle \tilde{\omega}_{+} = 0.0005$.
\bibitem{footnote:CubicLattice}
We verified that Z-point in our coordinate system is in the same direction as the K-point in the conventional cubic reciprocal lattice. 
The modulus of the Z-point is $|\Gamma Z| = \sqrt{2} (2 \pi)/a$, whereas the modulus of the K-point is $|\Gamma K| = \sqrt{2} (3 \pi)/(2a)$, corresponding to $\frac{3}{4}$ of $\Gamma Z$, in good mutual agreement.
\end{thebibliography}
\end{document}